\documentclass[preprint,prX]{revtex4}
\usepackage{graphicx,wrapfig}
\usepackage{amsfonts}
\usepackage{amsmath}
\usepackage{amssymb}

\def\bec{\begin{center}}
\def\eec{\end{center}}
\def\beq{\begin{equation}}
\def\eeq{\end{equation}}
\def\benu{\begin{enumerate}}
\def\eenu{\end{enumerate}}
\def\bit{\begin{itemize}}
\def\eit{\end{itemize}}
\def\beqr{\begin{eqnarray}}
\def\eeqr{\end{eqnarray}}
\def\beqrs{\begin{eqnarray*}}
\def\eeqrs{\end{eqnarray*}}
\def\btab{\begin{tabbing}}
\def\etab{\end{tabbing}}
\def\btable{\begin{tabular}}
\def\etable{\end{tabular}}

\def\om{\omega}
\def\gm{\gamma}
\def\Gm{\Gamma}

\def\lm{\lambda}

\def\dl{\delta}
\def\Dl{\Delta}
\def\sg{\sigma}
\def\Om{\Omega}

\def\del{\partial}

\def\bt{\beta}

\def\half{\frac{1}{2}}

\def\noi{\noindent}

\begin{document}

\title{Longitudinal Stability of Recycler Bunches \\
Part I: Thresholds for Loss of Landau Damping}

\author{T.~Sen}
\author{C.M.~Bhat}
\author{J.-F.~Ostiguy}
\affiliation{Fermi National Accelerator Laboratory \\ Batavia, IL 60510}

\begin{abstract}
We examine the stability of intense flat bunches in barrier buckets used
in the Recycler. We consider some common stationary distributions and 
show that they would be unstable against rigid dipole oscillations.
We then discuss an analytical model for the line density that best
describes measured bunch profiles. We include space charge in 
this model to predict the bunch intensity at which Landau damping 
would be lost. The dependence of this threshold on the bunch length is
studied and related to the results of an experimental study with shorter 
bunch lengths. The threshold for the microwave instability is estimated.
These studies will be followed by more detailed numerical studies.
\end{abstract}

\maketitle

\tableofcontents

\section{Introduction}

The creation of long flat bunches is under study for the LHC upgrade
as a way of increasing the luminosity \cite{Gonzalez}. The stability 
of such bunches
is one of the key issues of interest. At the Fermilab Recycler 
\cite{GJackson96}, long
flat bunches are created using rf barriers. At present
intensities these bunches are observed to be stable with lifetimes
around 20-50 hours (depending on intensity).
When electron cooling is enabled, the lifetime reaches $\sim 500$ hours 
at bunch intensities below 4.5$\times 10^{12}$. We explore the
longitudinal stability of these bunches at higher intensities
\cite{Griffin, Bhat06}. 

Bunches are confined within a barrier bucket by two voltage pulses of
equal magnitude and opposite polarity. The pulses, of equal duration $T_1$,
are separated by a duration $T_2$ with no applied voltage. The body of the 
bunch is contained within the interval $T_2$ but the head and tail of 
the bunch penetrate into the barrier on either side. 
Between the barrier pulses, the beam particles feel no longitudinal
focusing force and can be considered to be coasting. 
Changing $T_2$ adiabatically changes the 
bunch length while preserving the bunch area. The main longitudinal 
parameters of the Recycler bunches are
shown in Table \ref{table: param}. Single particle dynamics within 
such a bucket was studied in \cite{LeeNg}. Collective effects were
numerically studied in  the context of a low energy heavy ion ring \cite{OBF03}.
Longitudinal stability in the Recycler may be influenced by several
factors. For example, the synchrotron period
is rather long, hence even slowly growing instabilities which normally
be would not be of concern can in this case be important.
Furthermore, since the bunches are long, they can
be excited by relatively low frequency excitations.
\begin{table}
\bec
\btable{|c|c|c|} \hline
Parameter & Value & Units \\ \hline
Circumference & 3331.0        & m  \\
Energy & 8.93          & GeV \\
Bunch intensity & 4.5$\times 10^{12}$ & \\
$\gamma_t$ & 19.97         &    \\
Rf voltage $V_0$ & 1.8            & kV   \\
$T_1$ & 48.0    & 0.0189$\mu$ sec   \\
$T_2$ & 324.0   & 0.0189$\mu$ sec  \\
Bunch area & 100.0     & eV-sec  \\
$\nu_s^{max}$ & 9.97$\times 10^{-6}$  &   \\
\hline
\etable
\eec
\caption{Recycler bunch parameters}
\label{table: param}
\end{table}

In this paper we will consider primarily the thresholds for the
loss of Landau damping and also discuss the results of experiments
to determine the impact of a local instability possibly induced at
short bunch lengths. In Section II we discuss single particle dynamics
in a barrier bucket for the sake of completeness and handy reference.
In Section III we consider two stationary distributions
that are widely used to describe bunches confined by conventional harmonic 
rf systems as candidates to describe the Recycler bunches. 
In Section IV we 
calculate the coherent frequency of rigid dipole oscillations for
these distributions and the relationship to the maximum of the bare 
incoherent frequency.
In Section V we find an analytical distribution that adequately describes
the measured profiles. This distribution is then used to 
calculate the incoherent frequency distribution in the presence of
space charge. The intensity threshold at which Landau damping would be lost 
is predicted while keeping other
bunch parameters unchanged. In Section VI we consider the experimental
results obtained when the bunch length is changed and discuss the
dependence of the threshold for loss of Landau damping.
Section VII has a brief discussion of the microwave instability
and a rough estimate of the threshold intensity. We conclude in Section
VIII. In a subsequent paper we plan to a) discuss the stability diagrams
for the Recycler bunches and b) use numerical simulations to study 
the stability limits in greater detail.

\section{Single particle dynamics}

As mentioned above, a
barrier bucket consists of two voltage pulses of equal and opposite 
polarity, each lasting for time $T_1$ and separated by a time $T_2$.
The pulses can be of arbitrary shape, rectangular, triangular, half-sinusoidal
etc. In most of this report we will
consider a rectangular pulse profile.

Let $\tau$ be the time delay between an off-energy particle and the
synchronous particle which arrives at the center of the bucket $\tau=0$.
The equations of motion are 
\beqr
\frac{d}{dt}\tau & = & - \eta \frac{\Dl E}{\bt^2 E_0} \nonumber \\
\frac{d}{dt} \Dl E & = & \frac{e V(\tau)}{T_0}
\label{eq:eqmotion}
\eeqr
Here $\eta$ is the slip factor, and $T_0$ the revolution period. Since the 
bucket is centered at $\tau=0$, there is no change of energy for
$-T_2/2 \le \tau \le T_2/2$. The second order equation of motion for $\tau$ is
\beq
\frac{d^2}{dt^2} \tau + \frac{\eta}{\bt^2 E_0 T_0} eV(\tau) = 0
\eeq
From the first order equations of motion it follows that the Hamiltonian is
\beq
H = -\half \frac{\eta}{\bt^2 E_0} (\Dl E)^2 - \frac{e}{T_0}\int_0^{\tau}
V(s) ds = -\frac{\eta}{2\bt^2 E_0} (\Dl E)^2 - \frac{e}{T_0}U(\tau)
\eeq
where $U$ is the potential function.

At the bucket boundary or separatrix, the bucket height in energy $\Dl E_{bucket}$
is determined by
\beq
H_{bucket} = -\half \frac{\eta}{\bt^2 E_0} (\Dl E_{bucket})^2 = 
\frac{e}{T_0}\int_{T_2/2}^{T_2/2 + T_1} V(s) ds
\eeq
Hence the bucket height is 
\beq
\Dl E_{bucket} = \left[ \frac{2\bt^2 E_0}{T_0}
\left| \frac{1}{\eta}\int_{T_2/2}^{T_2/2 + T_1} eV(s) ds \right| \right]^{1/2}
\eeq

On any other level curve inside the boundary, the maximum time extent of
a particle is $T_2/2 + W$, where the parameter $W$ is the barrier penetration
depth for that particle. Then the maximum energy extent of a particle as
a function of $W$ is 
\beq
\Dl \hat{E} = \left[ \frac{2\bt^2 E_0}{T_0}
\left| \frac{1}{\eta}\int_{T_2/2}^{T_2/2 + W} eV(s) ds \right| \right]^{1/2}
\eeq
The particle energy deviation stays constant at this value for 
$-T_2/2 \le \tau \le T_2/2$.
At other times, the energy deviation can be found from
\beq
\Dl E(\tau)  =  \left[ (\Dl\hat{E})^2 - \frac{2\bt^2 E_0}
{\left|\eta \right| T_0}\int_{T_2/2}^{T_2/2 + W} eV(s) ds \right]^{1/2} 
 =   \left[ \frac{2\bt^2 E_0}{\left|\eta\right| T_0} 
\int_{T_2/2 + \tau}^{T_2/2+W} eV(s) ds\right]^{1/2}
\label{eq: DeltaE_tau}
\eeq
From the first of Equations (\ref{eq:eqmotion}) it follows that the time period on a phase curve is 
\beq
T_s  =  \frac{\bt^2 E_0}{\left|\eta\right|}[ 2\frac{T_2}{\Dl\hat{E}} + 4
\int_0^W\frac{d\tau}{\Dl E(\tau)} ] 
\label{eq: periodT_s}
\eeq
where $\Dl E$ in the integral is determined from the expression in 
Equation (\ref{eq: DeltaE_tau}).

We now specialize to the case of rectangular pulse profiles, i.e. the
voltage is given by 
\beqr
V(\tau) & = & -V_0  \;\;\;\;\;\;\;\; -(\half T_2 + T_1) \le \tau \le -\half T_2 
 \nonumber \\
 & = & 0 \;\;\;\;\;\;\;\;\;\;\;\; -\half T_2  \le \tau \le \half T_2 \\
 & = & V_0  \;\;\;\;\;\;\;\;\;\;\;\; \half T_2  \le \tau \le \half T_2 + T_1 \nonumber
\label{eq: rect_V0}
\eeqr
The potential which is only
determined up to an additive constant can be written as
\beqr
U(\tau) & = & -V_0[\tau + \half T_2 ] < 0 \;\;\;\;\;\;\;\; 
-(T_1 + \half T_2) \le \tau \le -\half T_2   \nonumber \\
 & = & 0 \;\;\;\;\;\;\;\;\;\;\;\;\;\;\;\;\;\;\;\;\;\;\;\;\;\;\;\;\;\;\;
  - \half T_2 \le \tau \le \half T_2  \nonumber \\
 & = & V_0[\tau - \half T_2 ] > 0 \;\;\;\;\;\;\;\;\;\;\; 
 \half T_2 \le \tau \le \half T_2 + T_1
\label{eq: U_flat}
\eeqr
Here we have chosen the additive constant so that the potential also
vanishes where the voltage does. The potential is zero outside this range
defined above. 

For this profile, the peak energy deviation on a phase curve with 
barrier penetration depth $W$, the bucket height and the time period on an
orbit are
\beqr
\Dl \hat{E} & = & \left[ 2 \frac{\bt^2 E_0}{\left|\eta\right|T_0}eV_0 W\right]^{1/2}, 
\nonumber \\
\Dl E_{bucket} & = & \left[ 2 \frac{\bt^2 E_0}{\left|\eta\right|T_0}eV_0 T_1\right]^{1/2} 
\nonumber \\
T_s & = & 2 \frac{\bt^2 E_0}{\left|\eta\right|}\frac{T_2}{\Dl \hat{E}} + 4 
\frac{T_0}{eV_0}\Dl \hat{E}
\eeqr
Note that the period does not go to infinity on the separatrix,
as is the case for harmonic rf.

The peak energy offset, at which the period reaches a minimum, is 
\beq
\Dl \hat{E}|_{min T_s} = \left[\frac{\bt^2 E_0 eV_0}{2\left|\eta\right|}
\frac{T_2}{T_0}\right]^{1/2} = \sqrt{\frac{T_2}{4T_1}}{\Dl E_{bucket}}
\label{eq: DelE_minTs}
\eeq
This leads to the conclusions that (a) if $T_2 > 4 T_1$, the period
minimum occurs outside the bucket, i.e. the synchrotron period just
decreases monotonically inside the bucket and (b) if $T_2 < 4 T_1$,
there is a point where the period reaches a minimum and then increases
up to the bucket. The region of zero slope in the period is 
associated with local loss of Landau damping. This simple
stability criterion requires that the interval between pulses 
be greater than four times the pulse width. 

The synchrotron tune on an orbit and the maximum tune 
when it occurs inside the bucket are respectively
\beqr
\nu_s  & = & \frac{T_0}{T_s} = \left[2 \frac{\bt^2 E_0}{\left|\eta\right|}\frac{T_2}{\Dl \hat{E} T_0} + 4 
\frac{\Dl \hat{E}}{eV_0}\right]^{-1}
 = (\frac{\left|\eta\right| eV_0 T_0}{2\bt^2 E_0})^{1/2} 
\frac{\sqrt{W}}{T_2 + 4 W} \nonumber \\
\nu_{s,max} & =  & \left[ \frac{\left|\eta\right|eV_0}{32\bt^2 E_0} \frac{T_0}{T_2}\right]^{1/2}
\label{eq: nus_max}
\eeqr
Note that this maximum tune is independent of the pulse width $T_1$.

The peak energy $\Dl \hat{E}$ on a phase curve is related to the
barrier penetration depth $W$ by
\beq
\frac{\Dl\hat{E}}{\Dl E_{bucket}} = \sqrt{\frac{W}{T_1}}
\label{eq: DeltaE_W}
\eeq

The synchrotron tune on a phase curve is related to the maximum tune as 
\beq
\frac{\nu_s}{\nu_{s,max}} = 4 \frac{\kappa}{1+ 4\kappa^2}, \;\;\;
\kappa =  \sqrt{\frac{W}{T_2}} = \sqrt{\frac{T_1}{T_2}}
\frac{\Dl\hat{E}}{\Dl E_{bucket}}
\eeq
where we have defined a dimensionless parameter $\kappa$ to label a 
phase curve.

On the phase curve where the synchrotron tune is maximum, $\kappa = 1/2$. 
At the bucket boundary, it has its maximum value $\kappa_{bucket} = \sqrt{T_1/T_2}$.

\vspace{1em}

\noi \underline{Area under phase curve}

The phase curve for a rectangular pulse profile consists of straight lines
at $ \pm \Dl\hat{E}$ between $-T_2/2$ and $T_2/2$ closed by parabolic paths.
On these parabolic paths, the relation between the time delay 
coordinate and the energy deviation  is
\beq
 \tau = \frac{\eta T_0}{2\bt^2 E_0 eV_0}[(\Dl\hat{E})^2-(\Dl E)^2], 
\eeq
while the area under each parabolic path is
\beq
\int_{-\Dl\hat{E}}^{\Dl\hat{E}} \tau d(\Dl E) = \frac{2}{3}
\frac{\left|\eta\right| T_0}{\bt^2 E_0 eV_0} (\Dl\hat{E})^3 
\eeq
Hence the area under a phase curve is
\beq
A = 2 \Dl \hat{E} T_2 + \frac{4}{3}
\frac{\left|\eta\right| T_0}{\bt^2 E_0 eV_0} (\Dl\hat{E})^3 = 
2 \Dl \hat{E} T_2[ 1 + \frac{4}{3} \kappa^2 ]
\label{eq: Area}
\eeq
while the bucket area is 
\beq
A_{bucket} = 2 \Dl E_{bucket} T_2[ 1 + \frac{4}{3} \frac{T_1}{T_2} ]
= 2 \left[ 2 \frac{\bt^2 E_0}{\left|\eta\right|T_0}eV_0 T_1\right]^{1/2}
 T_2[ 1 + \frac{4}{3} \frac{T_1}{T_2} ]
\label{eq: A_buck}
\eeq
If the maximum energy spread of the bunch $\Dl \hat{E}_{bunch}$ is 
known, e.g. from a longitudinal
Schottky measurement, then the bunch area can be found from Equation 
(\ref{eq: Area}). Currently the energy spread and the bunch area
of the Recycler beam are measured using a 1.7GHz Schottky detector.
For a beam with energy spread less than $\pm$23 MeV, the Schottky
frequency spread is less than $\pm$40 kHz, Hence in practice, the area 
under the Schottky spectrum 
between -40 MHz and 40 MHz is calculated, and the 
energy spread corresponding to 95\% of this area is found. This value is 
used to calculate the bunch area which contains 95\% of the particles. 

\vspace{1em}
\noi \underline{Bunch length}

The bunch barrier penetration depth $W_b$
can also be determined from the maximum energy spread in the bunch
using Equation (\ref{eq: DeltaE_W}), i.e. 
$W_b = (\Dl \hat{E}_{bunch}/\Dl E_{bucket})^2 T_1$. The bunch length
is $\tau_b = T_2/2 + W_b$. These equations express the fact that 
larger initial energy spread in the bunch lead to deeper barrier
penetration.

\section{Stationary distribution in a barrier bucket}

The time evolution of the phase space density $\rho$ follows from
\beq
\frac{d\rho}{dt} = \frac{\del\rho}{\del t} + \{\rho,H\}
\eeq
If the density is not explicitly time dependent and it is a function
of the Hamiltonian $H$ so that the Poisson bracket $\{\rho,H\}$ 
vanishes, then the density is stationary.

Due to the symmetry of the barrier, the phase curves are also 
symmetrical, each phase curve extending from $-W - 0.5 T_2$ to 
$W + 0.5 T_2$ on the $\tau$ axis (this defines $W$) and from $-\Dl \hat{E}$
to $\Dl \hat{E}$ on the energy axis. The peak energy $\Dl \hat{E}$ on a 
phase curve can be found from
\beq
(\Dl \hat{E})^2 = \frac{2\bt^2 E_0 e}{\left|\eta\right| T_0}\int_{0.5T_2}^{0.5T_2+W}
V(s) ds = \frac{2\bt^2 E_0 e}{\left|\eta\right| T_0}[U(\half T_2+W) - U(\half T_2)]
\eeq
At the bunch boundary on the $\tau$ axis $\tau = \tau_b$ and $\Dl E=0$,
and the Hamiltonian takes on the value
\beq
H = H_b = -\frac{e}{T_0} U(\tau_b) \equiv  -\frac{e}{T_0} U_b
\eeq
written in terms of the potential function $U_b$ at $\tau=\tau_b$. 

\subsection{Binomial distribution}

Consider a general binomial distribution for the density
\beq
\rho(\Dl E, \tau) = c_1 [H_b - H]^p
\eeq
where $p$ is a real number and $c_1$ is the normalization
constant.

The density can be expanded to the form
\beqr
\rho(\Dl E, \tau) & = & c_1[\frac{\eta}{2\bt^2 E_0}]^p
[ \frac{2\bt^2 E_0 e}{\eta T_0}[U(\tau) - U_b] - \Dl E^2]^p \nonumber \\
& = &  c_1[\frac{\eta}{2\bt^2 E_0}]^p
[\Dl E_b(\tau)^2 - \Dl E^2]^p
\eeqr
where we have used the fact that at any point on the bunch boundary, the
energy deviation can be found from
\beq
\Dl E_b(\tau)^2 = \frac{2\bt^2 E_0 e}{\left|\eta\right| T_0}[U(\tau) - U_b]
\label{eq: E_b}
\eeq
The line density is obtained by projecting the phase space density on to the 
$\tau$ axis,
\beq
\lm(\tau) = \int_{-\Dl E_b(\tau)}^{\Dl E_b(\tau)} \rho(\Dl E, \tau) d(\Dl E)
 = 2 c_1[\frac{\eta}{2\bt^2 E_0}]^p\int_0^{\Dl E_b(\tau)}
[\Dl E_b(\tau)^2 - \Dl E^2]^p
\eeq
where in the second equality we have used the fact that the integrand is
even. Using the integration result
\[ \int_0^{\Dl E_b} [\Dl E_b^2 - \Dl E^2]^p d(\Dl E) = 
\frac{\Gm(p+1)}{\Gm(p+3/2)} [\Dl E_b(\tau)]^{2p+1}
\]
we obtain for the line density
\beq
\lm(\tau) = \lm_0 [U(\tau) - U_b]^{p+1/2}, \;\;\;\;
\lm_0 = N_b [\int_{-\tau_b}^{\tau_b}(U(\tau) - U_b)^{p+1/2}]^{-1}
\eeq
where we have absorbed the constant $c_1$ and the other parameters
into a new constant $\lm_0$. 

The  distortion of the potential due to external inductive impedances and 
space charge can be included with the concept of an effective impedance. 
The effective impedance at frequency $\om$ is defined as
\beq
\frac{Z_{eff}}{n} = i(\om_0 L - \frac{g_0 Z_0}{2\bt\gm^2}) \equiv 
i\om_0 L_{eff}
\eeq
where $n = \om/\om_0$ is the harmonic of the revolution frequency, $L$ 
is the inductive impedance of the machine, $Z_0 = 377 \Om$ is the vacuum
impedance. The geometric space charge factor $g_0$ for a
round beam in a round beam pipe is given by
\beq
g_0 = \half + 2\ln (b/a)
\eeq
where $b$ is the radius of the beam-pipe and $a$ is the beam radius.

The induced voltage due to this impedance is 
\beq
V_{ind} = - L_{eff} \frac{dI}{dt} = - L_{eff} \frac{dI}{d\tau}
\eeq
From the bunch current $I(\tau) = e \lm(\tau)$, it follows that for a
binomial distribution,
\beq
\frac{dI}{d\tau} = e\lm_0(p+\half)[U(\tau) - U_b]^{p-\half}V(\tau)
\eeq
Introducing the notation
\beq
u(-\tau_b,\tau_b) = \int_{-\tau_b}^{\tau_b}[U(\tau) - U_b]^{p+\half} d\tau
\eeq
the induced voltage can be written as
\beq
V_{ind} = -\frac{e N_b(p+\half)}{\om_0 u(-\tau_b,\tau_b)} 
{\rm Im}[\frac{Z_{eff}}{n}]
[U(\tau) - U_b]^{p-\half}V(\tau)
\eeq

\subsection{Elliptic distribution}

The elliptic distribution is obtained by choosing $p = 1/2$ which yields
\beq
\lm(\tau) = \lm_0 [U(\tau) - U_b] 
\eeq
As we will see, this distribution has some special properties, first
pointed out by Hofmann and Pedersen \cite{HoffPed}.

\begin{figure}
\centering
\includegraphics[width=10cm]{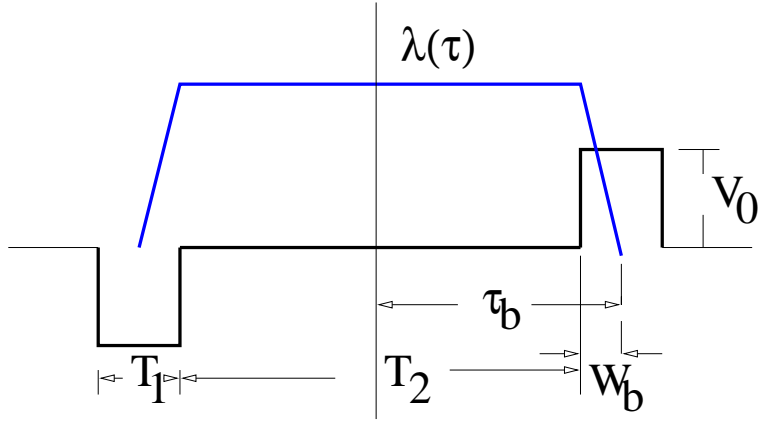}
\caption{Rectangular barrier bucket voltage profile and the line density
(in blue) in a stationary elliptic distribution. $T_1$ is the width of each
pulse, $T_2$ is the duration between the pulses, $\tau_b$ is half the bunch
length and $W_b$ is the barrier penetration depth.}
\label{fig: line_elliptic}
\end{figure}

The line density for an elliptic distribution with boundaries at 
$(-\tau_b,\tau_b)$ is
\beq
\lm(\tau) = \lm_0[U(\tau) - U(\tau_b)] = \frac{N_b}{u(-\tau_b,\tau_b)}
[U(\tau) - U(\tau_b)]
\eeq
Hence the stationary elliptic distribution in this rectangular barrier
bucket is
\beqr
\lm(\tau) & = & \frac{N_b}{[\tau_b^2 - (\half T_2)^2]}(\tau + \tau_b)
\;\;\;\;\;\;\;\; -\tau_b \le \tau \le -\half T_2 \nonumber \\
 & = & \frac{N_b}{[\tau_b^2- (\half T_2)^2]}(\tau_b - \half T_2)
\;\;\;\; -\half T_2 \le \tau \le \half T_2 \nonumber \\
& = & -\frac{N_b}{[\tau_b^2 - (\half T_2)^2]}(\tau - \tau_b)
\;\;\;\;\;\;\;\; \half T_2 \le \tau \le \tau_b 
\eeqr
This has a trapezoidal shape as shown in Figure \ref{fig: line_elliptic}.

For an elliptic distribution the induced voltage is
\beq
V_{ind} = -\frac{e N_b}{\om_0 u(-\tau_b,\tau_b)} {\rm Im}[\frac{Z_{eff}}{n}]
V(\tau)
\eeq
In this case it has the same form as the rf voltage so the total voltage is 
related to the focusing rf voltage by a change of scale. The total voltage is
\beq
V_t(\tau) = V(\tau) + V_{ind}(\tau) = \left\{1 - \frac{e N_b}{\om_0 u(-\tau_b,\tau_b)}
{\rm Im}[\frac{Z_{eff}}{n} ] \right\} V(\tau)
\eeq

By definition of the voltage waveform, a synchronous particle sees no focusing
voltage. Hence the ratio of the total focusing voltage to the rf focusing
voltage is 
\beq
k_t = \frac{V_t}{V} = 1 - \frac{e N_b}{\om_0 u(-\tau_b,\tau_b)}
{\rm Im}[\frac{Z_{eff}}{n}]
\label{eq: kt}
\eeq

The maximum synchrotron tune in the presence of space charge and external
impedances is 
\beq
\nu_{s}^{max} = [ \frac{e V_t \left|\eta\right|}{32\bt^2 E_0} \frac{T_0}{T_2}]^{1/2}
 = \sqrt{k_t} \nu_{s,0}^{max}
\eeq
where $ \nu_{s,0}^{max}$ is the maximum tune in the absence of space charge
and external impedances. Similarly the area under a phase curve with
turning points $(\tau_1,\tau_2)$ on the $\tau$ axis is given by
\beq
A_t(\tau_1,\tau_2) = \sqrt{k_t} A(\tau_1,\tau_2)
\eeq
The area of the bucket is reduced by the factor $\sqrt{k_t^{bucket}}$
given by Equation (\ref{eq: kt}) with $\tau_b = 0.5 T_2 + T_1$. 
The maximum theoretical intensity that can be stored in the bucket corresponds
to the case where the space charge and external impedances distort the
potential sufficiently to reduce the bucket area to zero. In practice,
the limiting bunch intensity will be lower than this value, for example
when the reduced bucket area can just accomodate the longitudinal
emittance of bunches from the Booster.

\subsubsection{Energy distribution}

The energy distribution can be found by projecting the phase space density
on to the energy axis. Let $\mu(\Dl E)$ represent the energy distribution,
then
\beq
\mu(\Dl E) \equiv \int \rho(\Dl E,\tau) d\tau = c_1 
[\frac{\left|\eta\right|}{2\bt^2 E_0}]^{1/2}\int_{-\tau_b}^{\tau_b}
[\left|\Dl E_b^2(\tau) - \Dl E^2 \right|]^{1/2}
\eeq
Note that $\mu$ is even in the energy. The energy offset on the 
bunch boundary is defined in Equation (\ref{eq: E_b}).
Over the three regions we have
\beqr
\Dl E_b^2(\tau) & = &  - \Dl E_{bucket}^2 \frac{\tau+\tau_b}{T_1}
\;\;\;\;\;\;\;\; -\tau_b \le \tau \le -\half T_2 \nonumber \\
 & = & - \Dl E_{bucket}^2 \frac{\tau_b - \half T_2}{T_1}
\;\;\;\;\;\;\;\;  -\half T_2 \le \tau \le  \half T_2 \nonumber \\
& = &  \Dl E_{bucket}^2 \frac{\tau - \tau_b}{T_1}
\;\;\;\;\;\;\;\; \half T_2 \le \tau \le \tau_b 
\eeqr
Consequently
\beqr
\mu(\Dl E) & = & c_1 [\frac{eV_0}{T_0}]^{1/2}\left\{
\int_{-\tau_b}^{-T_2/2}
|\tau+\tau_b - T_1\frac{\Dl E^2}{\Dl E_{bucket}^2}|^{1/2} d\tau 
 + |\tau_b - \half T_2 - T_1\frac{\Dl E^2}{\Dl E_{bucket}^2}|^{1/2}T_2 
\right.\nonumber \\
& & \left. + \int_{T_2/2}^{\tau_b}
|-\tau+\tau_b - T_1\frac{\Dl E^2}{\Dl E_{bucket}^2}|^{1/2} d\tau 
\right\} \nonumber \\
& = & \!\! c_1[\frac{eV_0 T_1}{T_0}]^{1/2}
\left\{ T_2\left|\frac{\Dl E^2}{\Dl E_{bucket}^2}  - 
\frac{W_b}{T_1} \right|^{1/2}
- \frac{4}{3}T_1 \left[ \left|\frac{\Dl E}{\Dl E_{bucket}} \right|^3  \!\! -  \!\!
 \left| \frac{\Dl E^2}{\Dl E_{bucket}^2}  \! -  \!
\frac{W_b}{T_1} \right|^{3/2} \right] \right\} \nonumber \\
\mbox{} 
\eeqr
This has its maximum at $\Dl E = 0$ and vanishes at 
\beq
\Dl E_{max} = \pm \Dl E_{bucket}\left[b + \frac{1}{2a(2-3ab)}
[3a^2 b^2 - 1 + \sqrt{(3a^2 b^2 -1)^2 + 4 a^3 b^3(2-3ab)}]\right]^{1/2}
\eeq
where $a = (4/3)T_1/T_2, \;\;\; b = W_b/T_1$.
For the Recycler parameters shown in Table \ref{table: param}, this 
yields a vanishing
of the energy distribution at $\pm$7.87 MeV well below the bucket
height $\Dl E_{bucket}= 17.38$ MeV.

\begin{figure}
\centering
\includegraphics[width=10cm]{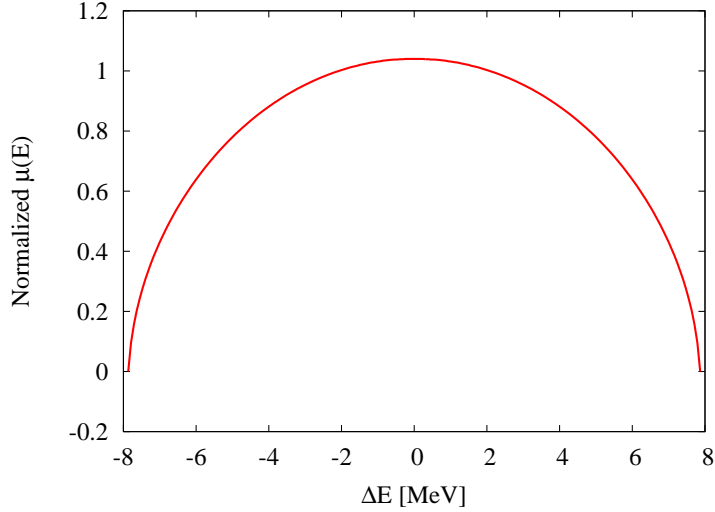}
\caption{The energy distribution as a function of the energy spread,
for the Recycler parameters, assuming an elliptic phase space 
distribution.}
\label{fig: energy_ellip}
\end{figure}

A sketch of the density distribution is shown in Figure 
\ref{fig: energy_ellip}, the parameters are those of the Recycler.
The plotted density is normalized to its maximum value at $\Dl E=0$. 
As is the case for the line density,
there is a well defined value beyond which the distribution $\mu(E)$ vanishes.

\subsection{Exponential distribution}

\begin{figure}
\centering
\includegraphics[width=10cm]{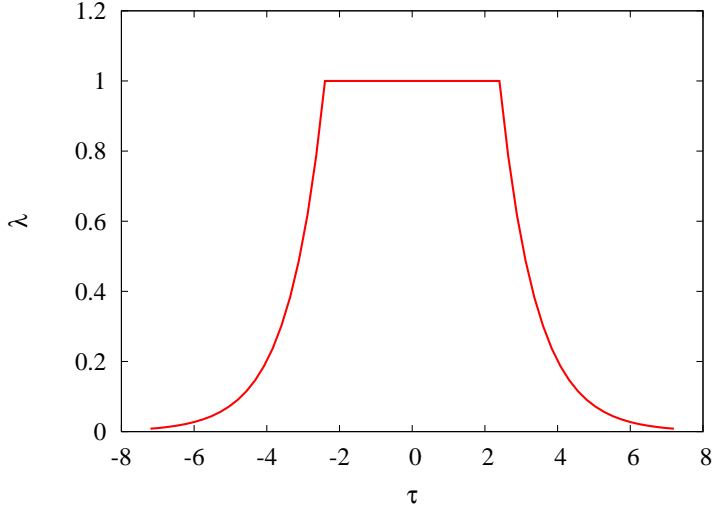}
\caption{Sketch of the line density for an exponential distribution.
Parameters are arbitrary.}
\label{fig: lambda_exp}
\end{figure}

The energy distribution for the elliptic phase space distribution does
not match the approximately Gaussian energy distribution that is observed in the
longitudinal Schottky spectrum of Recycler bunches. We therefore
consider another distribution which naturally leads to a Gaussian
energy distribution. 
Consider the density to be an exponential function of the Hamiltonian.
\beq
\rho(H) = \rho_0 \exp[ H/H_0] = \rho_0 
\exp\{\frac{1}{H_0}[\frac{-\eta}{2\bt^2 E_0}(\Dl E)^2 - \frac{eU(\tau)}{T_0}]
\}
\eeq
The line density obtained by projecting the density onto the time axis is
\beq
\lm(\tau) = \int \rho(\Dl E,\tau) d\Dl E = \lm_0 
\exp[ -\frac{e U(\tau)}{H_0 T_0}]
\eeq
where $\lm_0$ is a normalization constant and $H_0$ is a scale constant
to be defined later. From the expression in Eq (\ref{eq: U_flat}) for the 
potential function in a rectangular barrier bucket, it follows that
the line density for this distribution is 
\beqr
\lm(\tau) & = & \lm_0 \exp[\frac{e V_0}{H_0 T_0}(\tau + \half T_2 + T_1)]
\;\;\;\;\;\;\;\; 
-(T_1 + \half T_2) \le \tau \le -\half T_2   \nonumber \\
 & = & \lm_0 \exp[\frac{e V_0}{H_0 T_0}T_1]  \;\;\;\;\;\;\;\; 
\;\;\;\;\;\;\;\; \;\;\;\;\;\;\;\;\;
- \half T_2 \le \tau \le \half T_2   \nonumber \\
& = & \lm_0 \exp[-\frac{e V_0}{H_0 T_0}(\tau - \half T_2 - T_1)]
\;\;\;\;\;\;\;\; 
\half T_2 \le \tau \le \half T_2 + T_1
\label{eq: lm_exp}
\eeqr
Figure \ref{fig: lambda_exp} shows a sketch of the distribution.

Note that the density does not exactly vanish at $\tau_b$, instead
$
\lm(-\tau_b) = \lm(\tau_b) = \lm_0 \exp[-(e V_0[T_1-W_b ])/(H_0 T_0)]
$
where as before $W_b = \tau_b - \half T_2$. If $e V_0 [T_1-W_b]/(H_0 T_0) \gg 1$,
then the density is small and can be effectively taken to vanish at
$\tau=\tau_b$.

Since the density does not exactly vanish at $\pm \tau_b$, it would be
more accurate to extend the region of non-zero density to the entire
extent of the barrier bucket. In that case the 
normalization condition is $\int_{-(T_1+\half T_2)}^{T_1+\half T_2} 
\lm(\tau)d\tau = N_b$ which yields
\beq
 N_b = \lm_0 \left[
\frac{2H_0 T_0}{e V_0}(1 - \exp[-\frac{e V_0 T_1}{H_0 T_0}])  + T_2 \right]
\label{eq: norm_exp}
\eeq
If the line density is known in the flat region, e.g. at the center
$\tau=0$, then from the expression for the line density in
 Eq (\ref{eq: lm_exp}) above, it follows that the normalization constant
is known, i.e. $\lm_0 = \lm(\tau=0)$ and the normalization condition can
be used to find the scale constant $H_0$. It is simpler here
to determine this constant from the energy distribution which
in this case is a Gaussian
\beq
\mu(\Dl E) \equiv \int d\tau \rho(\Dl E, \tau) = 
\mu_0 \exp[-\frac{(\Dl E)^2}{2\sg_E^2}]
\eeq
where the rms energy deviation $\sg_E^2 = \bt^2 E_0 H_0/\eta$. If 
the rms energy deviation is measured from the Schottky spectrum, then
$H_0$ is known.

The line density written above is not self-consistent as it does not
take into account the potential well distortion due to space charge.
Including this intensity dependent effect, the line density is
\beqr
\lm(\tau) & = & \lm_0 \exp[-\frac{e U_t(\tau)}{H_0 T_0}] \\
U_t(\tau) & = & U_{rf}(\tau) +\int_0^{\tau} V_{ind}(s) ds 
   =  U_{rf}(\tau) + \frac{e}{\om_0}{\rm Im}[\frac{Z_{eff}}{n}]
         [\lm(0) - \lm(\tau)] \nonumber 
\eeqr
It follows that the density at time $\tau=0$ is 
$\lm(0) = \lm_0 \exp[-eU_{rf}(0)/(H_0 T_0)]$. Since we have defined
the rf potential such that $U_{rf}(0) = 0$, it follows that 
$\lm(0) = \lm_0$. Hence the self-consistent density is the solution of
the following equation
\beq
\lm(\tau) \exp[-\frac{e^2}{2\pi H_0}{\rm Im}[\frac{Z_{eff}}{n}]
\lm(\tau)] = \lm_0 \exp[-\frac{e^2}{2\pi H_0}
{\rm Im}[\frac{Z_{eff}}{n}]\lm_0]
 \exp[-\frac{e U_{rf}(\tau)}{H_0 T_0}]
\label{eq: Hassinski_exp}
\eeq
This is the equivalent of the Hassinski equation which defines a
self-consistent stationary solution for a sinusoidal rf and an 
arbitrary wake field. This equation can be solved numerically to
find the self-consistent density at a given intensity. At low
intensities this solution will reduce to the form  in Equation
(\ref{eq: lm_exp}).

\section{Frequency of coherent motion}

Imagine that the bunch is displaced from its center by $\Dl\tau$
as a result of which it starts to perform small amplitude oscillations. If
we let $\bar{\lm}(\tau)$ be the line density of the displaced bunch, then
it follows that $\bar{\lm}(\tau) = \lm(\tau-\Dl\tau)$. The infinitesimal
force on a slice of thickness $\dl\tau$ is 
\beq
dF = - \frac{e}{\bt c} \lm(\tau-\Dl \tau) d\tau \frac{\del U}{\del \tau}
\eeq
The total force on the bunch is
\beqr
F & = & - \frac{e}{\bt c} \int_{-\tau_b+\Dl\tau}^{\tau_b+\Dl\tau}
 \lm(\tau-\Dl \tau) \frac{\del U}{\del \tau} d\tau  \nonumber \\
 & = & \frac{e}{\bt c} \Dl\tau \int_{-\tau_b}^{\tau_b}
 \lm'(\tau) \frac{\del U}{\del \tau} d\tau + O[(\Dl\tau)^2] 
\eeqr
where we have expanded the displaced density to first order in $\Dl\tau$
and used the fact that the average force on the undisplaced bunch vanishes.
From the equation of motion and the force law 
$F \equiv - eV(\tau)/(\bt c) = m_{eff} \ddot{\tau}$,
it follows that the effective mass of a particle is 
$m_{eff} = \bt E_0 T_0/(c \eta)$ and hence the effective mass of the
bunch is $m_{eff}^{bunch} = N_b \bt E_0 T_0/(c \eta)$. If the centroid
is oscillating at a frequency $\om_c$, then from the small amplitude
equation of motion $F/m_{eff}^{bunch} \equiv \ddot{\tau_c} = -\om_c^2\Dl\tau$,
it follows that the coherent frequency is
\beq
\om_c^2 = \frac{e \left|\eta\right|}{\bt^2 E_0 T_0 N_b} \int_{-\tau_b}^{\tau_b}
\lm'(\tau) \frac{\del U}{\del \tau} d\tau
\label{eq: omegac_all}
\eeq
This coherent frequency is independent of intensity. It turns out
that for parabolic bunches in a single harmonic rf potential, the
coherent frequency and the small amplitude incoherent frequency
coincide. In general these frequencies are not the same for other
distributions and other rf systems.

\subsection{Landau damping for an elliptic distribution}

We will consider the rigid dipole mode to be Landau damped if the
frequency of the rigid dipole mode is within the band of incoherent
synchrotron frequencies and undamped if it lies outside this band.
First we evaluate the coherent dipole frequency for an elliptic
distribution. 
Using the line density of a matched binomial distribution, 
the coherent frequency using Eq. (\ref{eq: omegac_all}) is
\beq
\om_c^2 = \frac{e \left|\eta\right|(p+\half)}{\bt^2 E_0 T_0 u(-\tau_b,\tau_b)}
\int_{-\tau_b}^{\tau_b} [U(\tau)-U(\tau_b)]^{p-1/2}
(\frac{\del U}{\del \tau})^2 d\tau
\eeq
This frequency, in the absence of external forces, is independent of the
bunch intensity. For the elliptic distribution this reduces to 
\beq
\om_c^2 = \frac{e \left|\eta\right|}{\bt^2 E_0 T_0 u(-\tau_b,\tau_b)}
\int_{-\tau_b}^{\tau_b} V^2(\tau) d\tau   \label{eq: omegac}
= \frac{2 eV_0 \left|\eta\right|}{\bt^2 E_0 T_0 (\tau_b + \half T_2)}
\eeq

As remarked above, the total focusing voltage is changed by space charge
and external impedances. For an elliptic distribution, the total voltage 
is related to the rf voltage by the factor $k_t$ calculated above and the
synchrotron frequency is related to the bare synchrotron frequency by 
$\sqrt{k_t}$. In a barrier bucket, the incoherent frequency as a function
of amplitude rises from zero at the origin to a maximum value which may 
occur at an amplitude within the bucket. The maximum bare incoherent 
synchrotron angular frequency $\om_{s,0}^{max}$ can be found from
Equation (\ref{eq: nus_max}).
In the presence of space charge and external impedances it is modified
to 
\beq
\om_{s}^{max} = \sqrt{k_t}\om_{s,0}^{max} = 
  [1 + \frac{e N_b {\rm Im}(Z_{eff}/n)}{\om_0 V_0[\tau_b^2 - (\half T_2)^2]}]
^{1/2} \om_{s,0}^{max}
\eeq
In terms of the bare maximum frequency $ \om_{s,0}^{max}$, the coherent
frequency can be written as
\beq
\om_c = \frac{4}{\pi} [\frac{T_2}{\tau_b + \half T_2}]^{1/2}\om_{s,0}^{max}
\eeq
Writing $\tau_b = T_2/2 + W_b$, where $W_b$ can be considered as the
barrier penetration depth of the bunch, the requirement that the
coherent frequency be less than the maximum bare incoherent tune is
$W_b/T_2 \ge \frac{16}{\pi^2} - 1$.

Below transition with space charge as the dominant impedance, the
net focusing voltage is reduced and the incoherent frequency decreases with
intensity. At the threshold for the loss of Landau damping, the
maximum incoherent frequency equals the coherent dipole frequency and falls
below it at higher intensities. The threshold intensity for the loss of 
Landau damping found by equating $\om_{s}^{max}$ and $\om_c$ is 
\beq
N_b^{thresh} = - \frac{\om_0 V_0(\tau_b^2 - (\half T_2)^2)}{e {\rm Im}(Z_{eff}/n)}
[ 1 - (\frac{4}{\pi})^2 \frac{T_2}{\tau_b + \half T_2} ]
\eeq

Putting in numbers for the Recycler, we find that 
\beq
\frac{\om_c}{\om_{s}^{max}} = 1.25
\eeq
This says that even at zero intensity, the coherent frequency would
be outside the incoherent band. As the intensity increases, the 
incoherent frequencies will decrease and the ratio above will increase.
Hence there would be no Landau
damping at any intensity. We conclude that this distribution
cannot be suitable to describe the Recycler longitudinal density.


\subsection{Landau damping for an exponential distribution}

The coherent frequency can be found using Eq (\ref{eq: omegac_all}) except
that we now extend the region of integration so
\beq
\om_c^2 = \frac{e \left|\eta\right|}{\bt^2 E_0 T_0 N_b} 
\int_{-(\half T_2+T_1)}^{\half T_2+T_1}
\lm'(\tau) \frac{\del U}{\del \tau} d\tau
\eeq
We find 
\beq
\om_c^2 = \frac{2 e V_0 \left|\eta\right|}{\bt^2 E_0 T_0} 
\frac{1 - \exp[-\frac{e V_0 T_1}{H_0 T_0}]}
{2\frac{H_0 T_0}{e V_0}(1 - \exp[-\frac{e V_0 T_1}{H_0 T_0}]) + T_2}
\eeq
Expressed in terms of the maximum of the bare incoherent frequency
$\om_{s,0}^{max}$,
\beq
\frac{\omega_c}{\omega_{s0,max}} = \frac{4}{\pi}
[ \frac{1 - \exp(-\chi W_b)}{1 + (2/(\chi T_2))
[1 - \exp(-\chi W_b)]}]^{1/2}, \;\;\;\;
\chi = \beta^2 E_0 eV_0/(\left|\eta\right|\sigma_E^2 T_0)
\eeq
Putting in the Recycler numbers we find that in this case
\beq
\frac{\om_c}{\om_{s0,max}} = 1.27
\eeq
This distribution is also unsuitable.

\section{Measured distributions}

During a study in April 2009, 
protons were injected into the Recycler with a increasing number of
Booster batches. The bunch profiles were measured with a wall current
monitor at three different intensities. Longitudinal Schottky spectra
were also recorded. The measured bunch length profiles 
are well described by the function 
\beqr
\lambda_{fit}(\tau) & = & N_b a\left\{ [1+{\rm tanh}(b\tau+c)]\theta(-\tau)
 + [1 - {\rm tanh}(b\tau -c)]\theta(\tau) \right\} \\
\lm_{fit}(0) & = & N_b a(1 + {\rm tanh}c)   \nonumber
\eeqr
Here $\theta$ is the step function, $a, b, c$ are the three parameters
of the fit. $a$ sets the overall normalization for the density, $b$ is a
measure of the slope in the rising and falling regions and 
the width of each half section is directly proportional to $c$.
Note that the same parameters describe both the head and
the tail of the profile, implying perfect symmetry about the center.

\begin{figure} 
\centering
\includegraphics{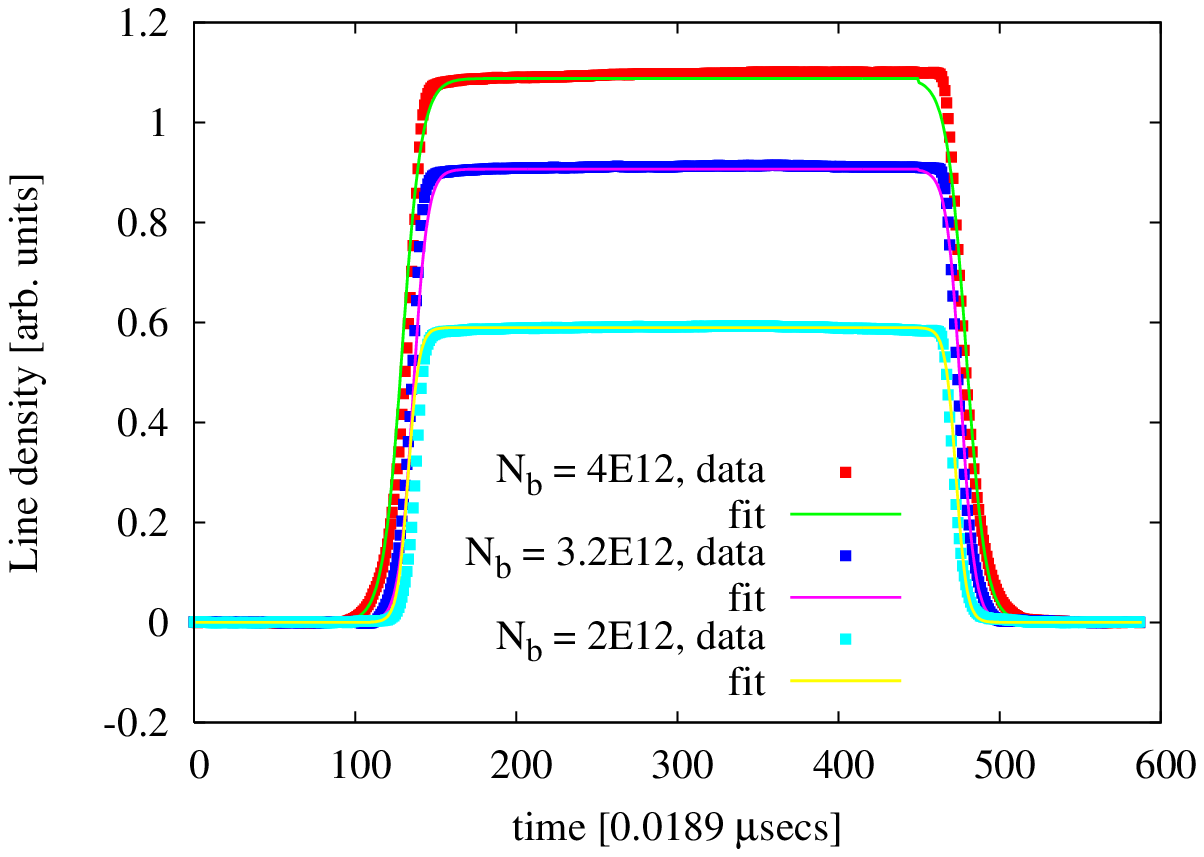}
\caption{Comparison of the measured line density and fit to the
data with a tanh profile at three different intensities .}
\label{fig: full_fit_all}
\includegraphics{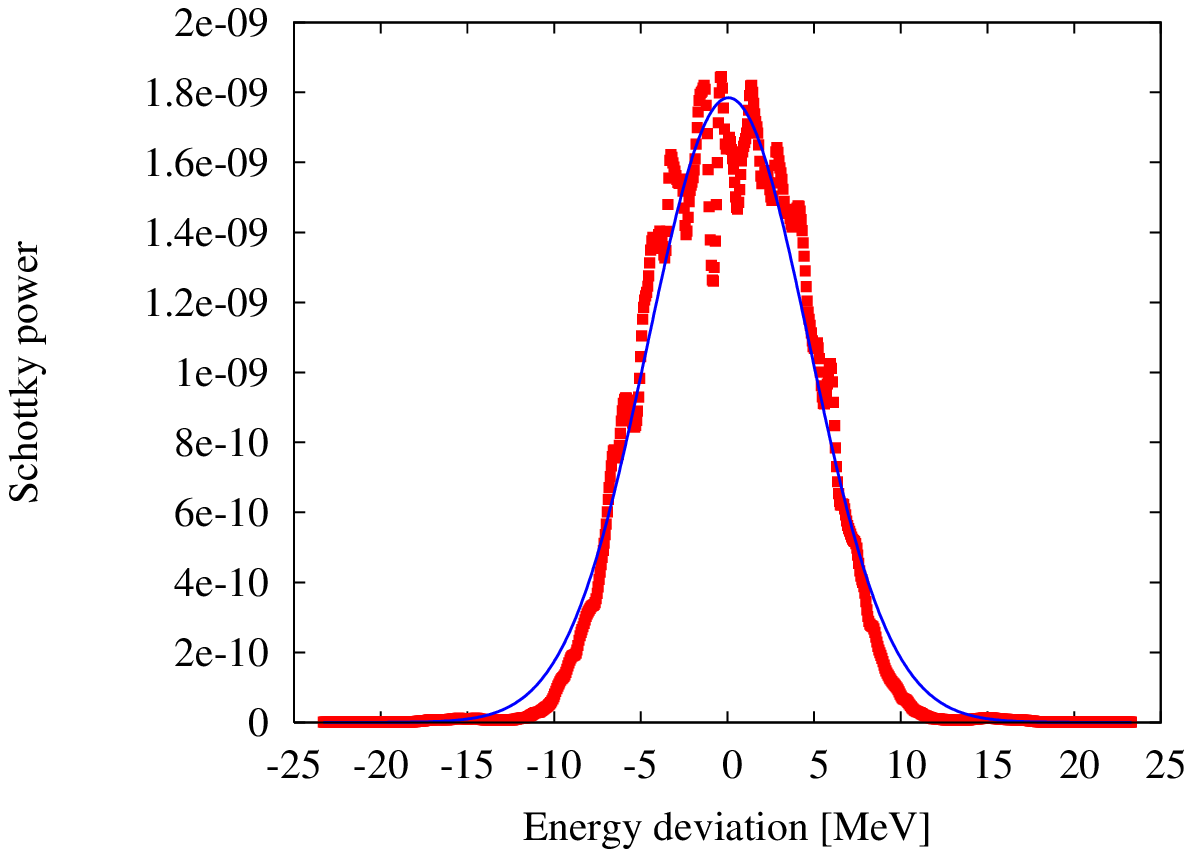}
\caption{Measured Schottky power of a proton beam in the Recycler
compared with a Gaussian fit.}
\label{fig: Schottky_meas_09}
\end{figure}

The measured profiles and the fit function are shown for all three sets
in Figure \ref{fig: full_fit_all}. 
This fitted function appears to describe the measured profiles with
very good accuracy. Note that the same fit works equally well at
all the intensities suggesting that distortion due to intensity
effects are not yet important at these levels. This also suggests
that the $tanh$ fit is probably not the result of a self-consistent 
solution to the ``Hassinski equation'' (\ref{eq: Hassinski_exp}), the 
assumption being that the
fit is also good for lower intensities than those measured here.
There is some bunch
lengthening as the intensity increases showing the impact of the
capacitive space charge impedance. 

\begin{figure} 
\centering
\includegraphics{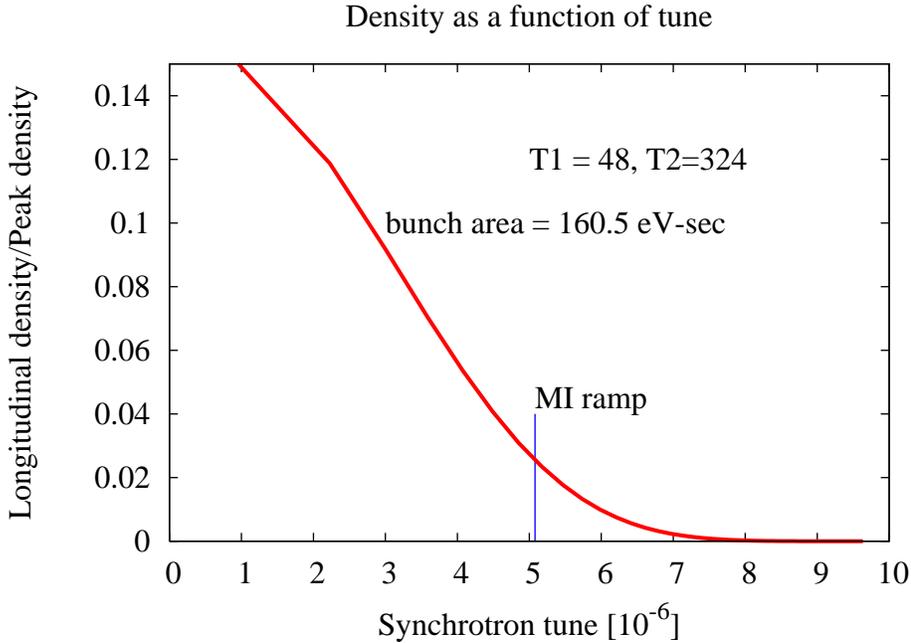}
\caption{The relative density as a function of the bare incoherent
tune. Most particles have zero tune. For comparison, the tune
corresponding to the frequency (0.45 Hz) of the Main Injector ramp
is also shown. }
\label{fig: lambda_nus}
\end{figure}

The line density can also be calculated as a function of the 
incoherent synchrotron tune. Figure \ref{fig: lambda_nus} shows
the density relative to the density at the center as a function of
$\nu_s^0$, the bare incoherent tune. The  Main Injector ramping
frequency is also shown in this figure; beam particles resonant with
this frequency can be driven to larger amplitudes and lost. However the
relative density at this frequency is about 2\%, the net effect due to
this resonance should be small at present parameters. 

\begin{figure} 
\centering
\includegraphics{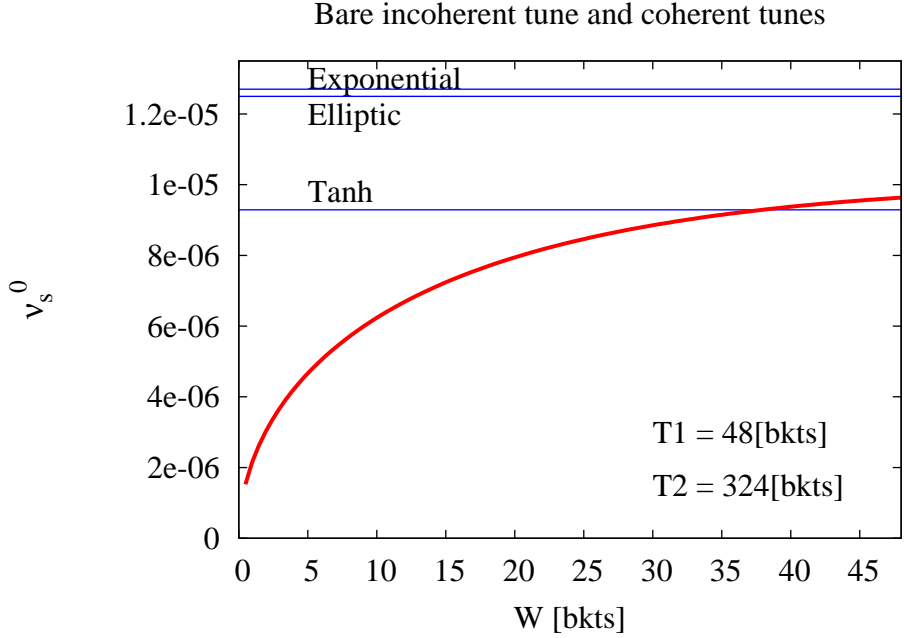}
\caption{Bare synchrotron tune vs $W$, the barrier penetration
parameter for each orbit. Also shown are the coherent tunes of
the dipole mode for three distributions. The coherent mode would be
damped only for the Tanh distribution.}
\label{fig: coh_alldistrib}
\end{figure}

We can now use this form of the line density
to calculate the coherent frequency. From the expression for the
coherent frequency in Equation (\ref{eq: omegac_all})
\beqr
\om_c^2 & = & \frac{\left|\eta\right| eV_0}{\bt^2 E_0 T_0 N_b}
\{ - [\lm(-\half T_2) - \lm(-\tau_b)] + [\lm(\tau_b)-\lm(\half T_2)] \}
 \nonumber \\
 & = & \frac{2 \left|\eta\right| eV_0 a}{\bt^2 E_0 T_0 }
\{ {\rm tanh}(b\tau_b - c) - {\rm tanh}(\half b T_2 - c) \}
\eeqr
Hence the ratio of the coherent frequency to the maximum of the
bare incoherent frequency is
\beq
\frac{\omega_c}{\omega_{s0,max}} = \frac{4}{\pi}\sqrt{a T_2}
\sqrt{{\rm tanh}[b\tau_b-c]-{\rm tanh}[\half b T_2- c]}
\eeq
Here we use the value of the parameter $a$ determined by the normalization
condition, i.e. 
\beq
a = \half [ \half T_2 + T_1 + \frac{1}{b}\log(\frac{\cosh c}
{\cosh[b(T_2/2+T_1)-c]})]^{-1}
\eeq
Using the values of the fit parameters $b, c$ for the 
data measured in the April 2009 study we find that
\beq
\frac{\om_c}{\om_{s}^{max}} = 0.93
\eeq
i.e. at zero intensity the coherent frequency is within the 
incoherent spread. Figure \ref{fig: coh_alldistrib} shows the
coherent tune for the three distributions considered here
and the bare incoherent tune as a function of the parameter $W$.

\subsection{Threshold for loss of Landau damping}

We can now take into account the potential well distortion due to space
charge and find the incoherent tune in the presence of space charge.
The total voltage for an arbitrary distribution is 
\beq
V_t = V_0 - \frac{e}{\om_0}{\rm  Im}(\frac{Z_{eff}}{n})\frac{d\lm}{dt}
\eeq
The energy deviation on a curve with barrier penetration $W$ at time
 $\tau$ is
\beqr
(\Dl E)^2(\tau;W) \!\! & \!\!  = \!\!  & \!\! \frac{2\bt^2 E_0}{\left|\eta\right|T_0}
\int_{T_2/2 + \tau}^{T_2/2 + W} eV_t(s) ds  \nonumber \\
& \!\! = & \!\!\! \frac{2\bt^2 e E_0}{\left|\eta\right|T_0}\left\{ V_0(W-\tau) - 
 \! \frac{e}{\om_0}{\rm  Im}(\frac{Z_{eff}}{n})[\lm(\half T_2 + W)
 -  \!\! \lm(\half T_2 + \tau)] \right\}
\eeqr
Introducing the function $f(W)$ and the parameters $\zeta, d$ as
\beq
f(W) = W + \zeta {\rm tanh}(bW+d), \;\;\; 
\zeta = \frac{e}{\om_0 V_0}{\rm  Im}(\frac{Z_{eff}}{n})N_b a , \;\;\;
d = \half b T_2 - c
\eeq
we can write 
\beq
(\Dl E)^2(\tau;W) = \frac{2\bt^2 e V_0 E_0}{\left|\eta\right|T_0}[f(W) - f(\tau)]
\eeq
while the peak energy is found from $\Dl \hat{E} = \Dl E(0,W)$.
The intensity dependence is contained in the parameter $\zeta$.

\begin{figure} 
\centering
\includegraphics{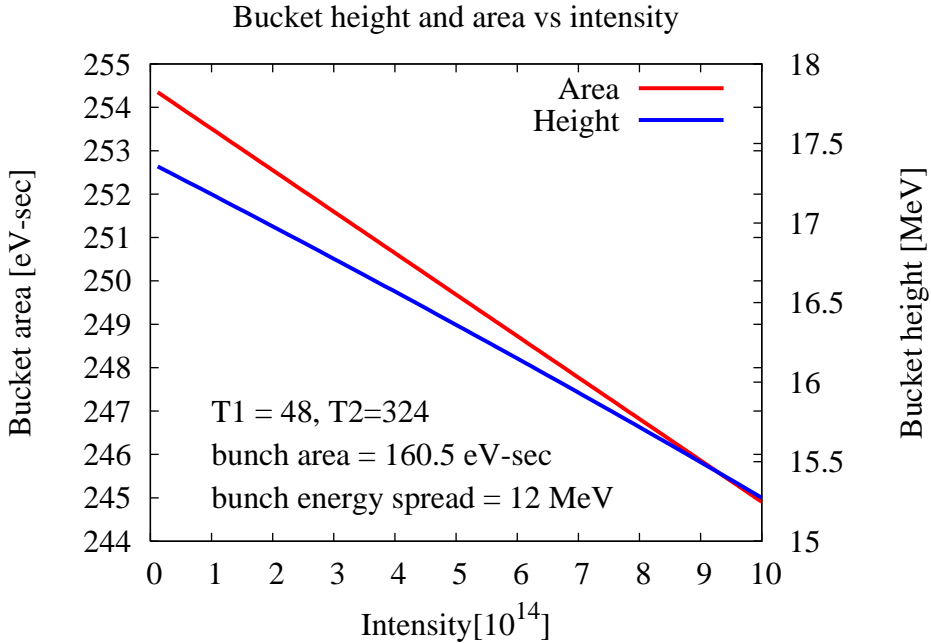}
\caption{The bucket area and bucket height as a function of the
bunch intensity. Over the range of intensities plotted, the 
bucket values of the area and height stay well above the initial 
bunch area and bunch energy spread. }
\label{fig: bucket_Nb}
\end{figure}

As the intensity increases, the bucket height and bucket area 
decrease. A threshold for the maximum intensity that can be 
stored in the bucket is set when the bucket area falls to the
initial bunch area. Figure \ref{fig: bucket_Nb} shows that at
intensities even up to 200 times larger than present, the bucket area
and height are larger than the initial bunch area and energy spread
respectively. The threshold is at an intensity of 2.3$\times 10^{15}$, where
the bucket height equals the initial bunch energy spread of 12 MeV.

The incoherent synchrotron tune as a function of $W$ can then be found 
using Eq (\ref{eq: periodT_s}) as
\beq
\nu_s(W) = (\frac{\left|\eta\right| eV_0 T_0}{2\bt^2 E_0})^{1/2}
\left[ \frac{T_2}{[f(W) - \zeta{\rm tanh}d]^{1/2}}
 + 2 \int_0^W \frac{d\tau}{[f(W) - \tau - \zeta{\rm tanh}(b\tau+d)]^{1/2}}
\right]^{-1}
\eeq
The integral in the above expression can be performed numerically.
For comparison, the bare synchrotron tune without space charge is
given in Equation (\ref{eq: nus_max}).

\begin{figure} 
\centering
\includegraphics{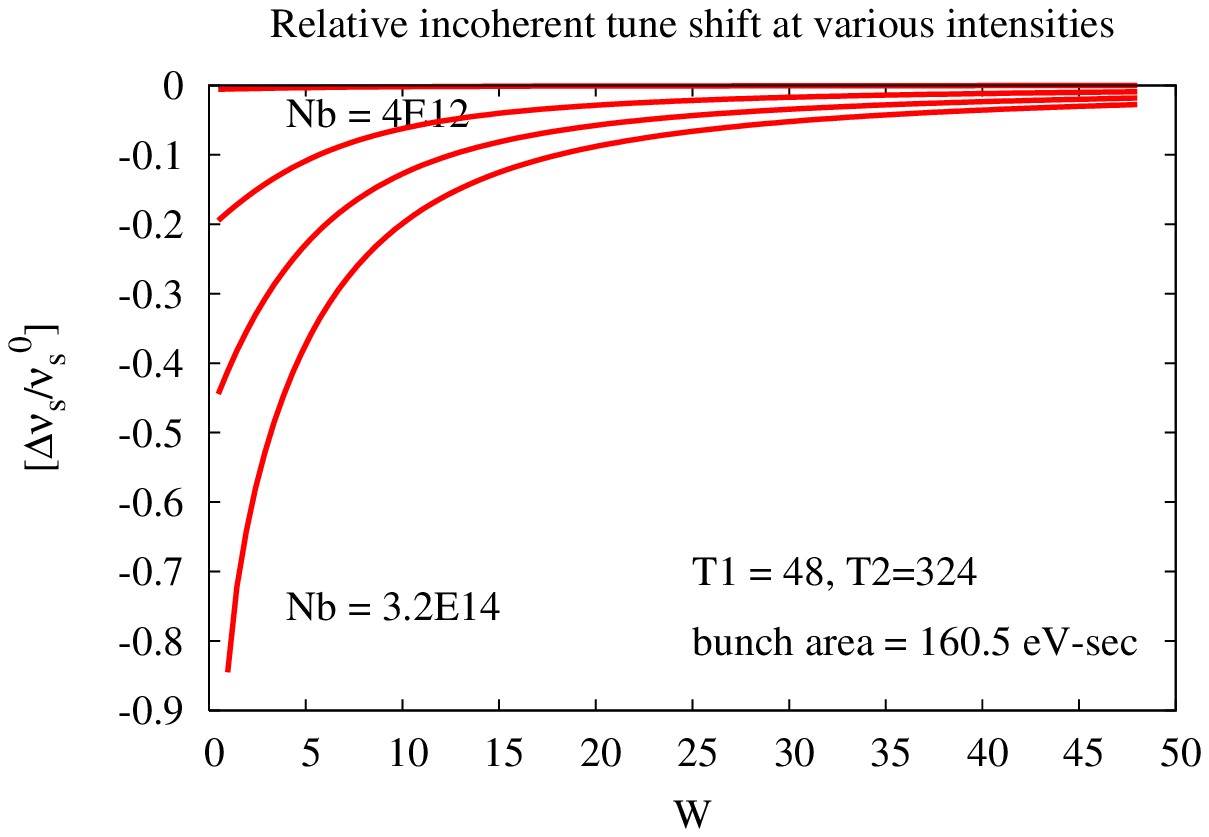}
\includegraphics{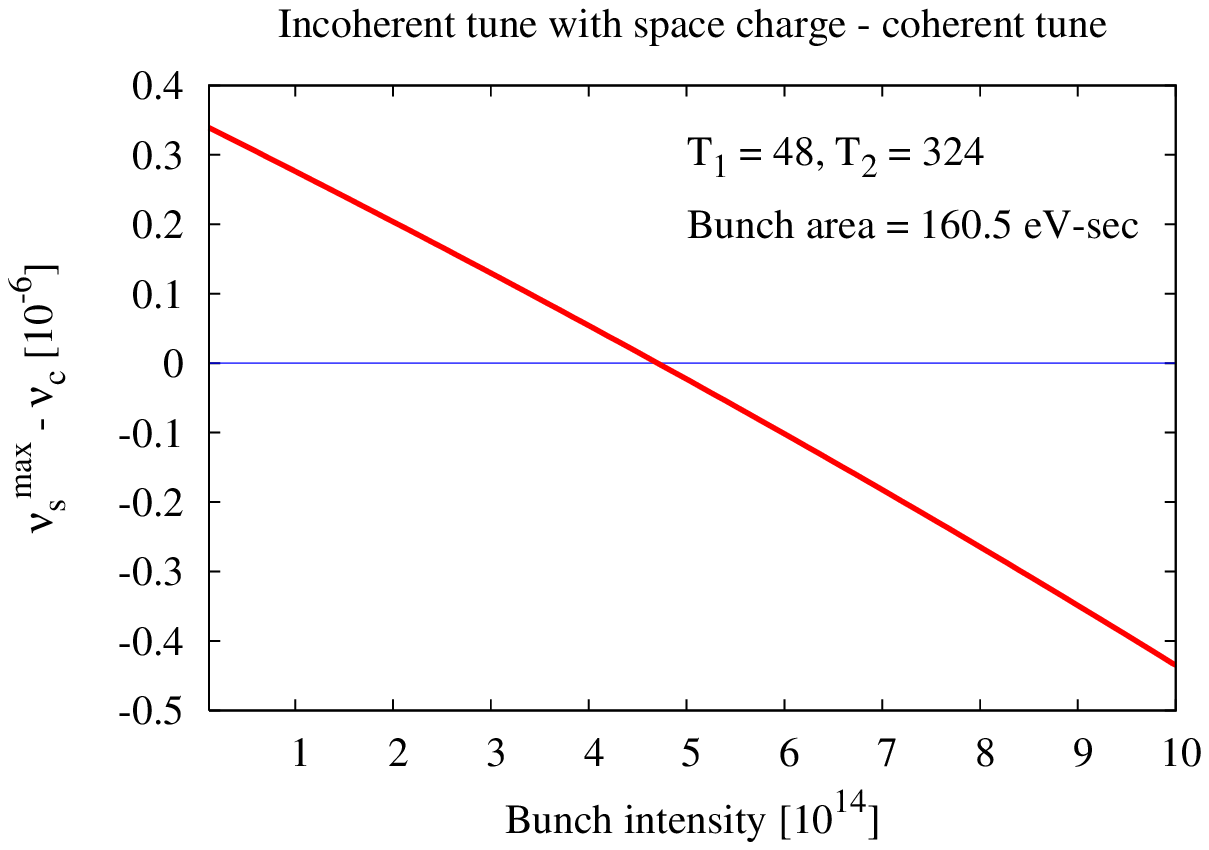}
\caption{Top: Relative difference of the incoherent tune with space 
charge and without space charge as a function of W, the barrier 
penetration at bunch intensities varying from $N_b = 4\times 10^{12}$ 
to $N_b = 3.2\times 10^{14}$.
Bottom: Difference of the maximum incoherent tune with space charge
and the coherent frequency  (multiplied by $10^6$) as a function of 
the bunch intensity. The zero crossing defines the threshold for the
loss of Landau damping, here the threshold is 
$N_b = 4.7\times 10^{14}$.}
\label{fig: incoh_coh_tunes}
\end{figure}

The top plot in Figure \ref{fig: incoh_coh_tunes} shows the relative 
difference of the incoherent tunes with and 
without space charge at intensities varying over two orders of
magnitude . The relative 
tune depression due to space charge is the largest at
the inner edge of the voltage pulse since the tune is also smallest
there. 
The incoherent frequency decreases as the intensity increases and
at a threshold frequency, the maximum of the incoherent tune will
will equal the coherent frequency. At higher intensities, the
coherent frequency will be outside the spread and Landau damping
is lost. For the nominal value of $T_2=324$ 'bkts', this threshold
is reached when the intensity exceeds 4.7$\times 10^{14}$,
about 100 times higher than present intensities.

\section{Instability thresholds at different bunch lengths}

It has been observed in the SPS that a region where the incoherent
frequency as a function of amplitude has an extremum is associated with
the appearance of a local instability \cite{Shaposhnikova}. 
In a barrier bucket, the maximum of the incoherent tune occurs when the
ratio of the peak energy on the orbit to the bucket height is
$\sqrt{T_2/(4T_1)}$, see Equation (\ref{eq: DelE_minTs}). If this ratio is 
greater than unity, as is the
case for the nominal Recycler parameters $T_2/(4T_1) = 324/(4\times 48)$,
the maximum lies outside
the bucket. The effects of this local instability are not present
under normal operation but can be studied by changing the separation
$T_2$. 

The energy acceptance or bucket height depends on the area under the
voltage pulse $V_0 T_1$ but not on $T_2$. 
If we change the bunch length by adiabatically changing $T_2$, the
bunch area will be preserved and the energy acceptance will also
be constant. The bucket area however is determined by the bucket height
and the value of $T_2$. The value of $T_2$ 
when the bucket area just encloses the bunch area is
\beq
T_2^{min} = \half\sqrt{\frac{\left|\eta\right|T_0}{2\bt^2 E_0 eV_0 T_1}}A_{bunch}
 - \frac{4}{3}T_1
\eeq
Lower bunch area leads to a lower value for $T_2^{min}$ and hence allows a 
wider range of variation with $T_2 < 4 T_1$. 

These studies were carried out in the Recycler using protons.
The beam cooling systems as well as the transverse dampers were 
turned off during the experiment. For these measurements, a bunch
with much lower intensity than normal was injected - the initial bunch area
was about 27 eV-sec compared to a more typical value of 100 eV-sec. The
bunch was injected into a gap which was 84 Booster buckets long, i.e.
$T_2 = 84$ 'bkts' (a 'bkt'$=0.01893\mu sec$). We chose $T1=48$ bkts 
 and pulse height of 1.8kV, these being the usual values during regular
operation. Once equilibrium was reached after injection, beam intensity 
along with  data from the wall current and Schottky
monitors were recorded to establish the initial
parameters. Subsequently the beam was expanded or compressed adiabatically 
by changing $T_2$ to different values of interest without changing $T_1$.

\begin{figure} 
\centering
\includegraphics[scale=0.55]{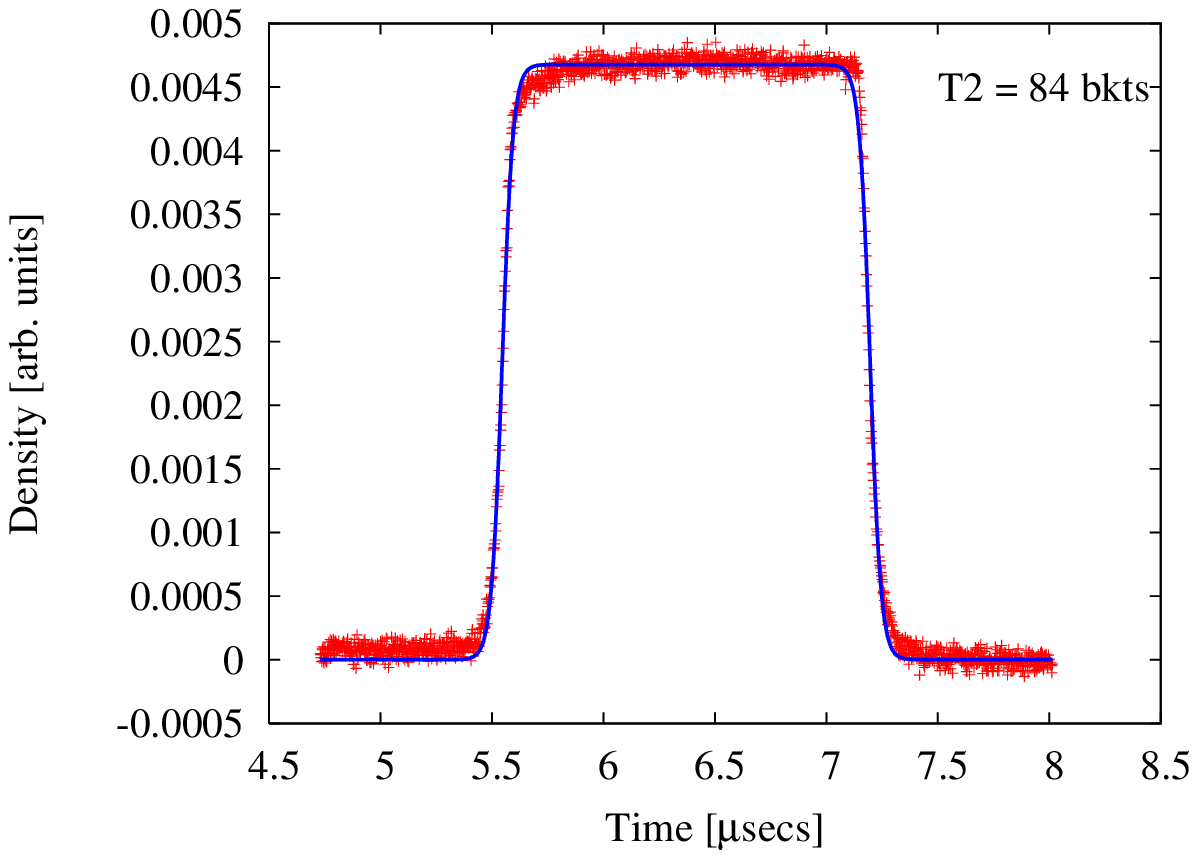}
\includegraphics[scale=0.55]{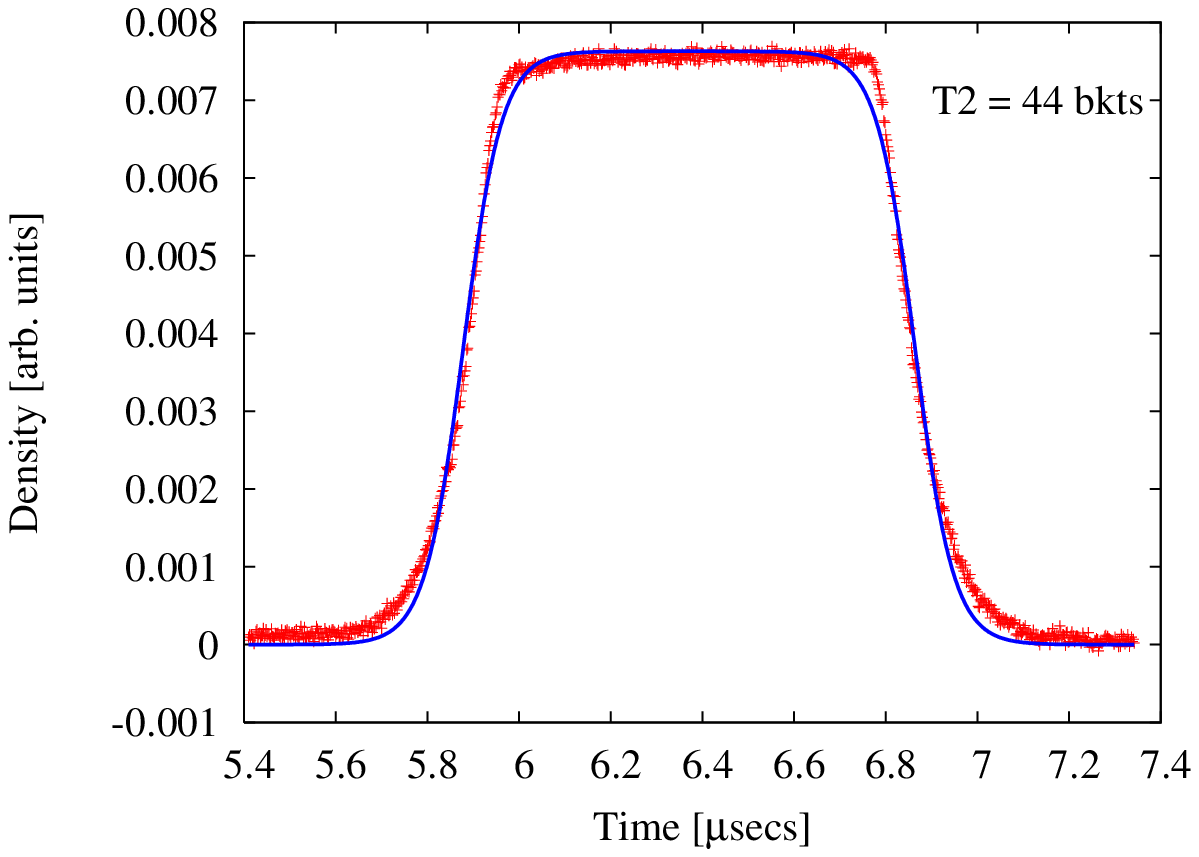}
\includegraphics[scale=0.55]{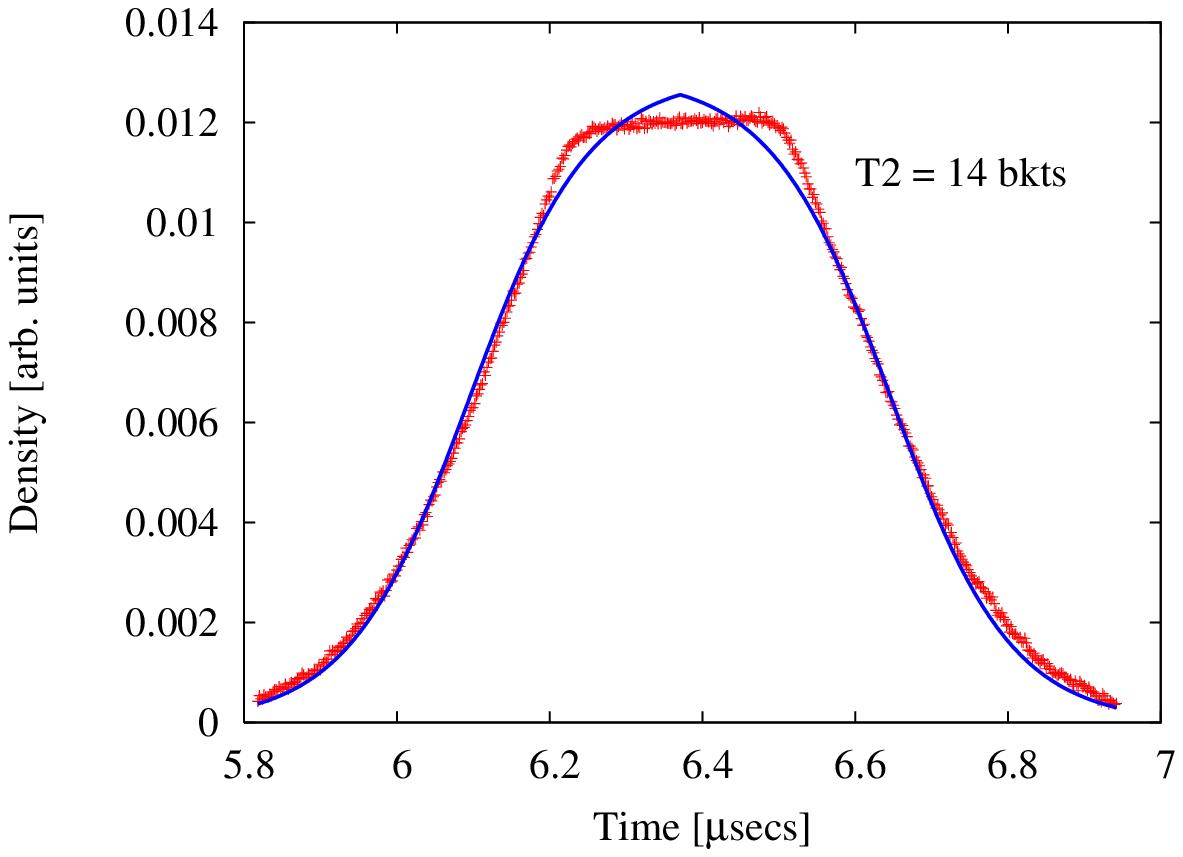}
\includegraphics[scale=0.55]{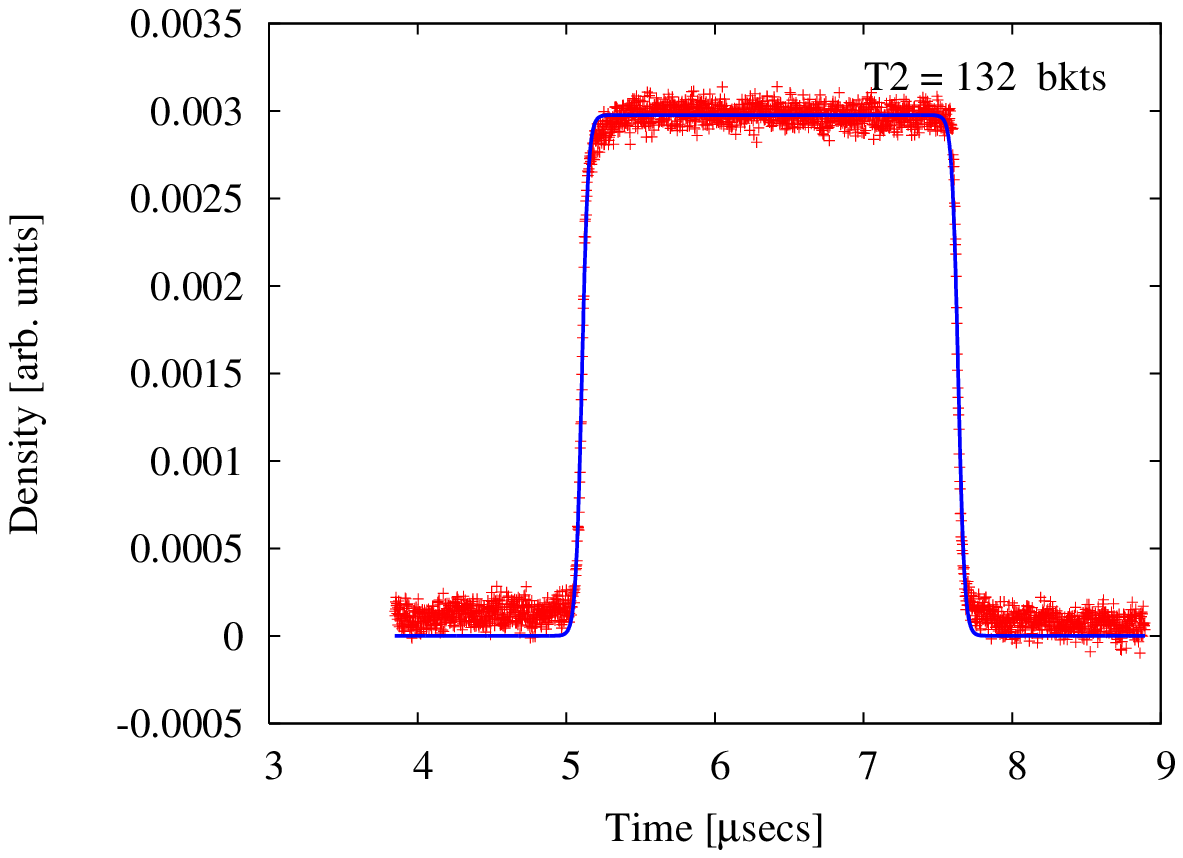}
\includegraphics[scale=0.55]{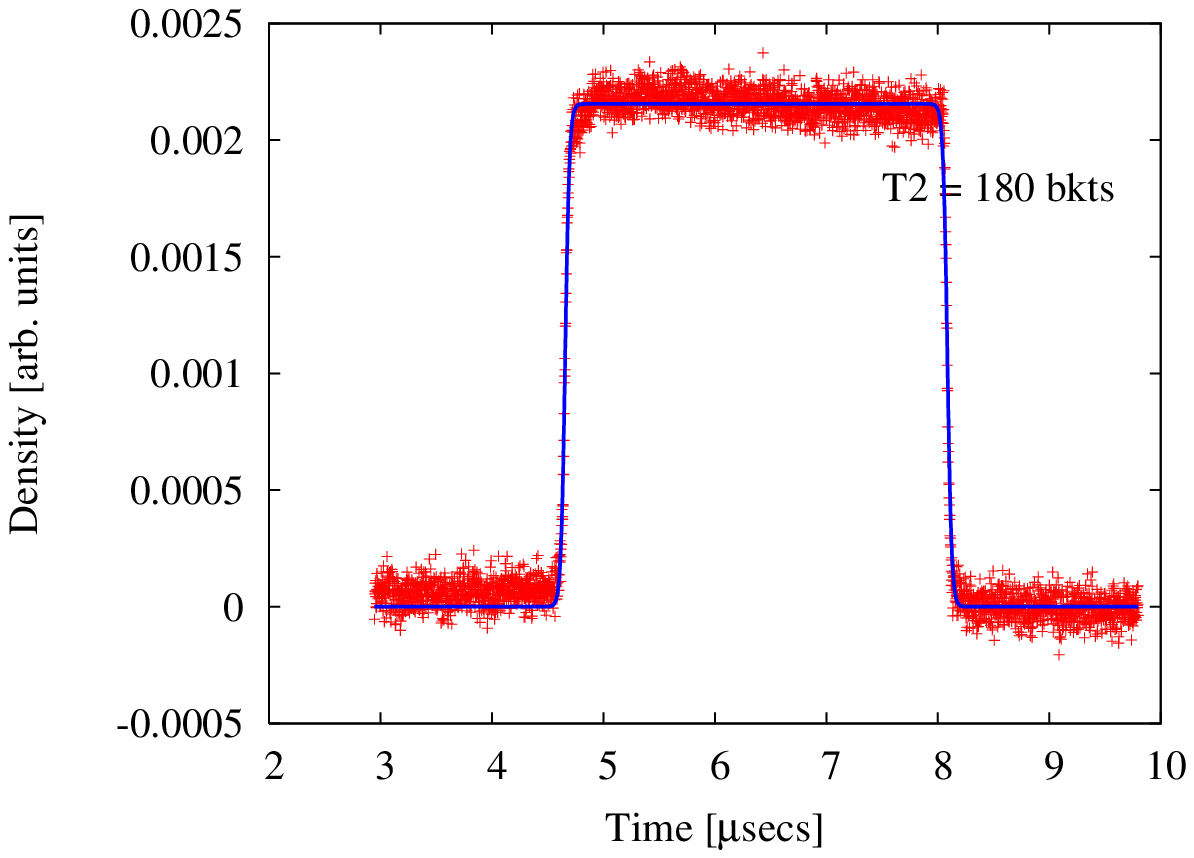}
\includegraphics[scale=0.55]{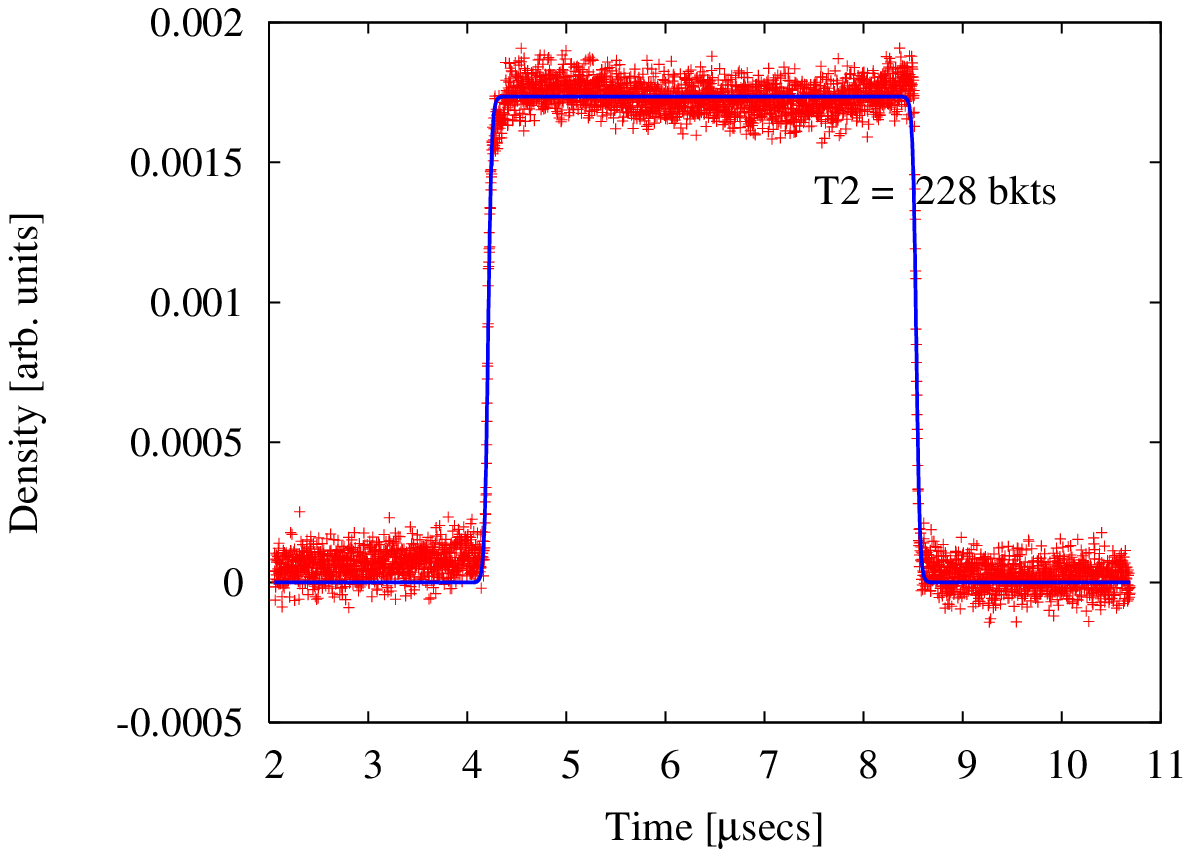}
\includegraphics[scale=0.55]{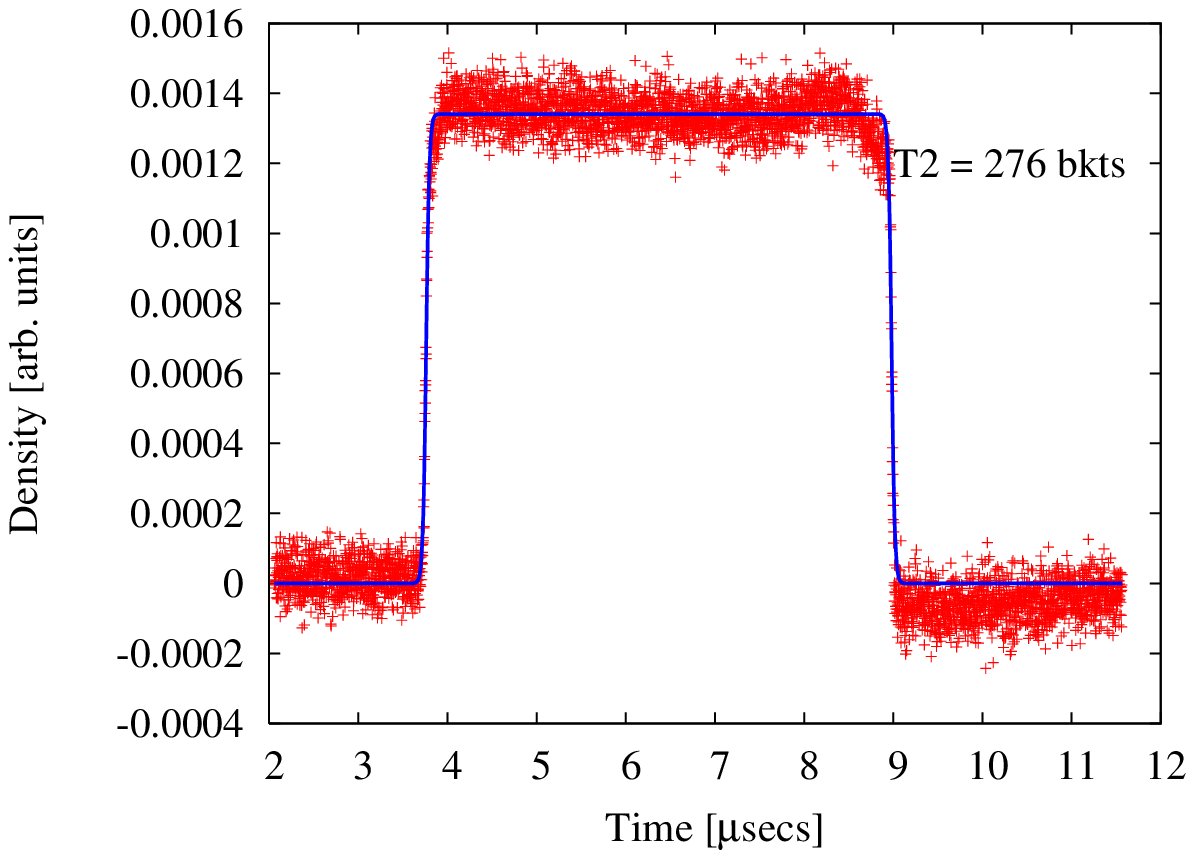}
\includegraphics[scale=0.55]{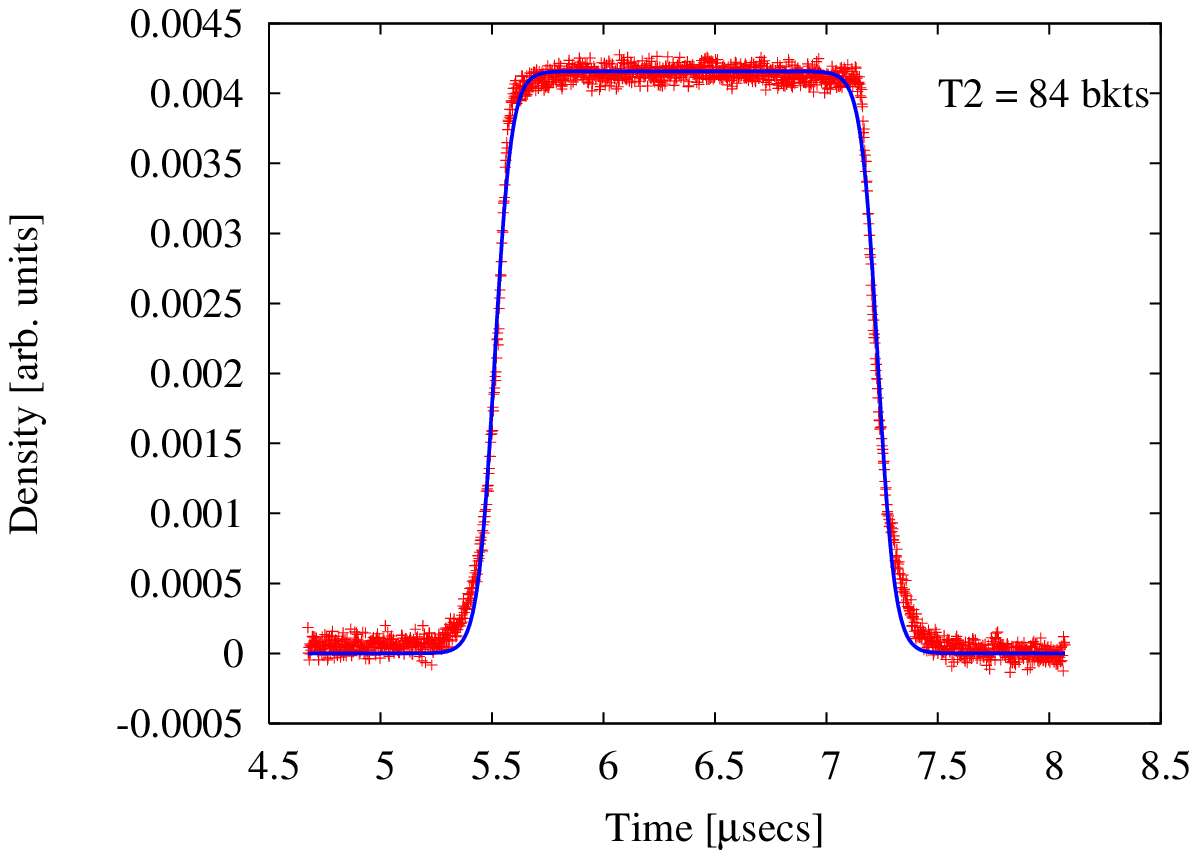}
\caption{Measured bunch profiles in March 2009. The value of $T_2$
was changed while $T_1$ was kept constant. The sequence of $T_2$ values
is the measurement order. The measured data and
the fits with the {\em tanh} function are shown for the eight
profiles.}
\label{fig: Mar09_lineprofiles}
\end{figure}
Figure \ref{fig: Mar09_lineprofiles} shows the measured longitudinal
profiles along with the fits by the {\em tanh} function. In all cases,
the fit describes the measured profile quite well. One slight exception
is for the shortest bunch with $T_2=14$ bkts where the fit at the center
does not quite match the flatness of the measured profile. 
Furthermore, it cannot be explained by the wall current monitor saturation
which occurs at a minimum of 5 times the bunch intensity used in this
experiment. However the
rising and falling edges match very well. These edges in the barrier
pulses determine the coherent frequency, not the bunch profile in the
center. 

\begin{figure}
\centering
\includegraphics[scale=0.8]{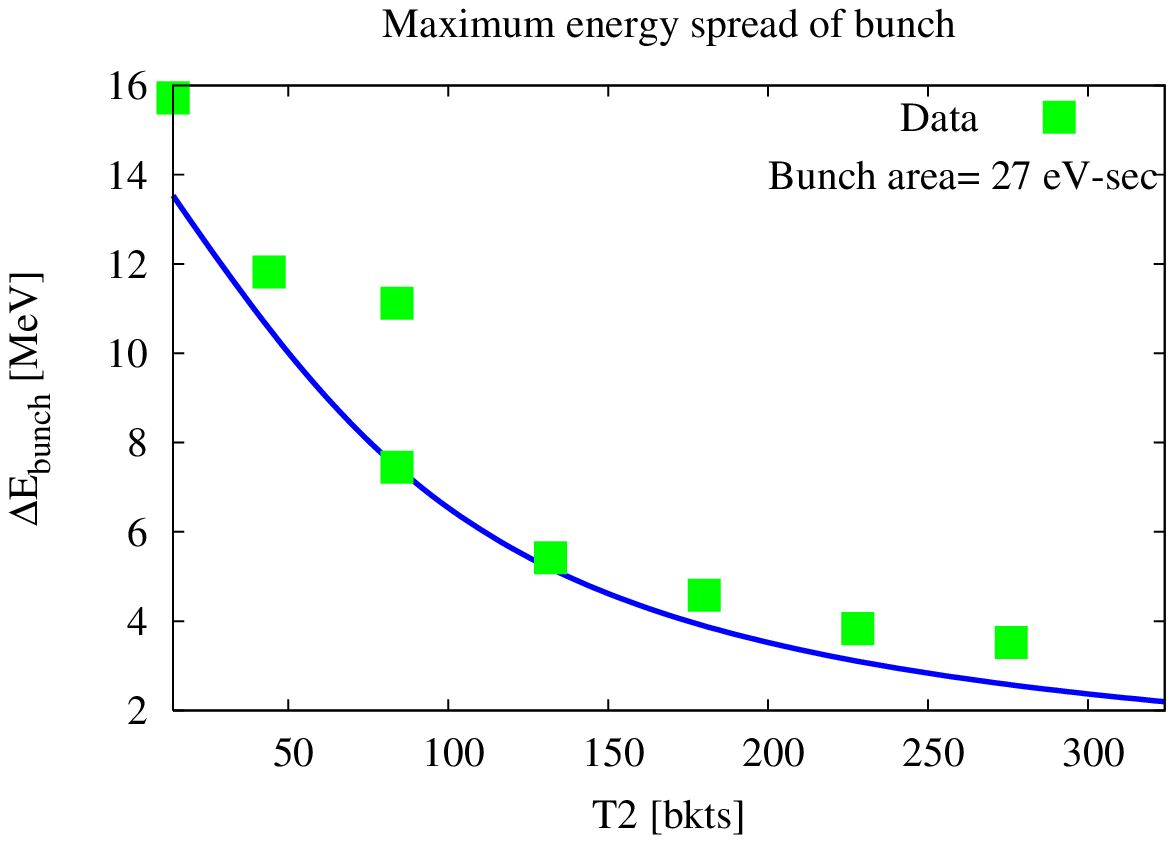}
\caption{The maximum energy spread as a function of $T_2$.
The data points are the values obtained from the Schottky spectrum
while the curve shows the expected value assuming
that the initial bunch area is preserved and no beam is lost. Over this 
range of $T_2$
from 14 bkts to 324 bkts, the maximum spread stays below the
energy acceptance of 17.3 MeV. The deviations between the data and
the curve show that the bunch area was not preserved, especially at
the smallest value of $T_2 = 14$bkts.}
\label{fig: DelE_T2}
\end{figure}

\begin{figure}
\centering
\includegraphics{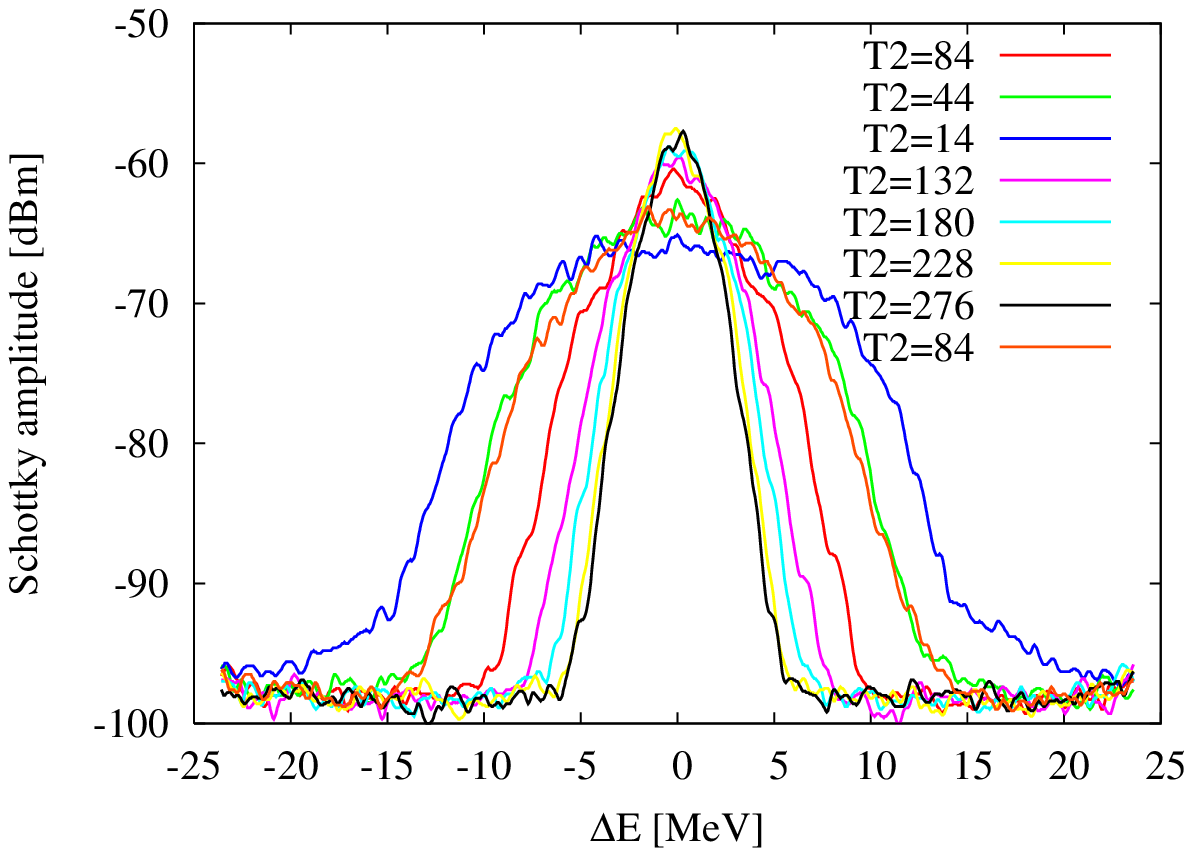}
\caption{Schottky spectra of bunches with different values of $T_2$ after
background subtraction.
At the smallest values of $T_2$, the spectra extends beyond the energy
acceptance suggesting that there was some beam loss but this amount was
of the order of 1\%.}
\label{fig: Schottky_T2}
\end{figure}

Assuming that the bunch area is preserved, compressing the bunch length
increases the energy spread and conversely. Figure \ref{fig: DelE_T2}
shows the maximum energy spread deduced from the Schottky spectrum
and the theoretical values (shown by the curve) assuming that the
initial bunch area was conserved. There are discrepancies between the
two, especially at the shortest $T_2=14$ bkts. Some of this can be 
attributed to measurement errors of the energy spread from the
Schottky spectrum. It is also likely that several
beam dynamics processes lead to an increase in bunch area and loss of
particles as the gap $T_2$ was decreased. This is confirmed by the
measured Schottky spectra for the different profiles shown in
Figure \ref{fig: Schottky_T2}. At the smallest values of $T_2$,
the energy spread exceeds the energy acceptance of 17.4 MeV.
However the spectral amplitudes, shown in db, also show that the
losses were of the order of 1\%.

\begin{figure}
\centering
\includegraphics[scale=0.8]{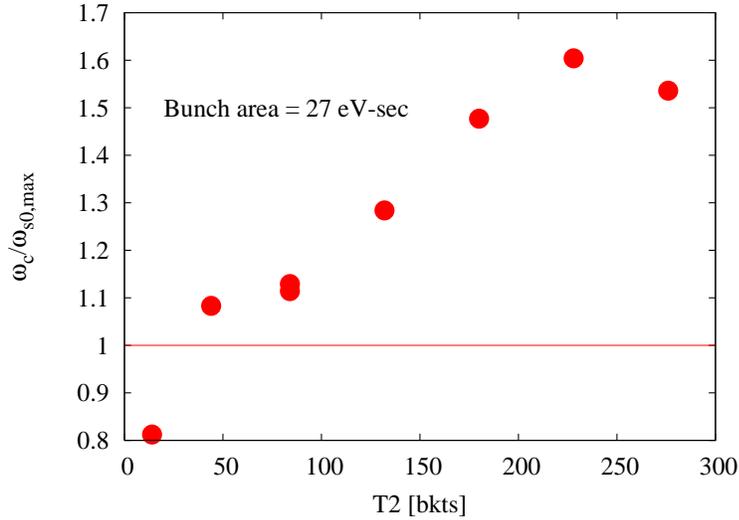}
\caption{Ratio of coherent to maximum of the bare incoherent frequency
for the different bunches with line profiles shown in Figure
\ref{fig: Mar09_lineprofiles}.}
\label{fig: Mar09_coh_incoh}
\end{figure}

We now consider the coherent frequency and the ratio of this frequency
to the maximum of the bare incoherent frequency as a function of the
bunch length, using the energy spread and the fitted values of the
parameters $b,c$ for each of the bunch profiles shown in Figure
\ref{fig: Mar09_lineprofiles}. This ratio is shown in Figure 
\ref{fig: Mar09_coh_incoh}.
First, the ratio increases with the bunch length or $T_2$ before falling 
at the largest value of $T_2$. Note that both the coherent frequency
and the maximum of the bare incoherent frequency {\em decrease} with $T_2$
. $\om_{s0,max}$ decreases as $1/\sqrt{T_2}$ while the coherent frequency
decreases at a slower rate, hence the ratio increases with $T_2$.
 One would conclude from this observation that
for bunches with the same (or nearly the same here) longitudinal 
emittance, shorter bunches have a higher threshold for the loss of Landau
damping. Secondly, we observe that except for $T_2=14$, this ratio is
greater than one implying that Landau damping was lost for most of the
bunches in this study. This may explain some of the observed increase in
bunch area. 

\begin{figure}[htb]
\centering
\includegraphics*[scale=0.5]{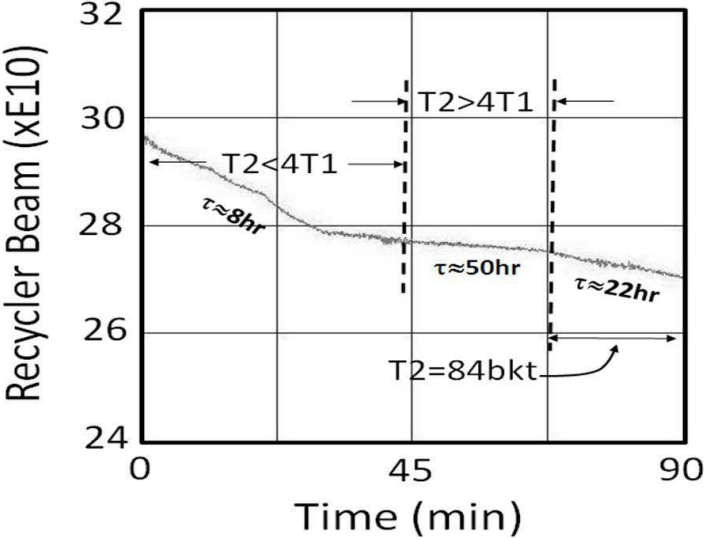}
\caption{Lifetime at various times with different $T_2$ to $T_1$ ratios. }
\label{fig: lifetimes}
\end{figure}

Fig.~\ref{fig: lifetimes} shows the surviving beam as a function of time.
Various $T_2$ to $T_1$ ratios along with the beam lifetime are also
indicated.
There are three distinct regions in this plot. The first section had
$T_2 < 4 T_1$ with $T_2$ assuming values of $14, 44, 84, 130, 180$ bkts.
 The average lifetime of the beam at 8 hours was significantly lower
than in the second section where it was about 50 hours with $T_2 =228, 276$ and
$>4T_1$ in both cases. In the third region the beam 
was compressed to its original value of $T_2 = 84$ bkts. The lifetime 
again decreased by about a factor of two. There are many possible
sources for these changes in lifetime as $T_2$ was changed: different
energy spread, loss of Landau damping, local instability when 
$T_2 < 4 T_1$, and possible resonance with the frequency of the Main
Injector ramping. The density of particles at this resonant frequency
changes with the bunch length and this resonance has been known to cause 
beam loss in the past. While we do not at present know the relative
importance of these effects, the data is
not inconsistent with the hypothesis that a local instability at
the extremum of the incoherent synchrotron tune can 
contribute to beam loss. 
The effects of this instability are easily avoided in the barrier bucket
by choosing appropriate values of $T_1, T_2$


\section{Microwave instability}

So far we've only discussed the effect of the capacitive space charge
impedance. However the Recycler also has a broadband resistive impedance
with a cutoff at $f_c \sim c/b \approx 3.95$ GHz. At sufficiently high
intensities, this resistive impedance can lead to asymmetric bunch shape 
distortion
and eventually to turbulent bunch lengthening due to the onset of the
microwave instability. While there is as yet no satisfactory theory of
this instability for bunched beams, estimates for the threshold of
the instability are made by using the coasting beam theory and replacing
the beam current by the peak bunch current. For Gaussian bunches, the
intensity threshold is given by 
\beq
N_{\mu} =\sqrt{\frac{\pi}{2}}
\frac{Z_0}{r_p} \frac{\left|\eta\right|\gamma}{\beta}\frac{\sigma_z \sigma_{\delta}^2}
{\left|Z_{||}/n \right| }
\eeq
We observe that for constant bunch area $A \sim \sg_z \sg_{\delta}$, the
threshold $N_{\mu} \propto A^2/\sg_z$. Hence for bunches of the same area,
shorter bunches have a higher threshold.

As a preliminary estimate for the threshold of this instability, we will
use the rms values of the energy spread and the bunch length for
the measured distribution during normal operation. The rms energy spread
is found from the Schottky spectrum while the rms bunch length can be 
found from
\[ \tau_{rms} = \sqrt{a}[\int_{-(T_2/2+T_1)}^0 \tau^2(1+{\rm tanh}(b\tau+c))d\tau + \int_0^{(T_2/2+T_1)} \tau^2(1-{\rm tanh}(b\tau-c))d\tau ]^{1/2}
\]
For the fitted values of the parameters, we find $\tau_{rms} = 3.095\mu$
seconds while the rms energy spread is $\sg_E = 4$MeV. 
The effective wall impedance is estimated to be $Z_{||}/n = 1.6$ Ohms.
Substituting
the values we find that the bunch intensity threshold for the microwave 
instability is
\beq
N_{\mu} = 2.95\times 10^{15}
\eeq
This is about three orders of magnitude above present intensities. 
Space charge below transition does not induce the microwave
instability but some results, e.g. \cite{OBF03} suggest that it may moderate
the growth of the instability. 
Simulations may be necessary to understand this process.

\section{Conclusions}

Our primary findings are as follows:
\bit
\item We considered two typical stationary phase space distributions
as possible candidates for the Recycler: an elliptic distribution
and an exponential (in the Hamiltonian) distribution. The first has
an energy distribution which does not match observations while the 
second has a longitudinal profile distribution which does not
match observations. In addition, both of these distributions are above 
the threshold for loss of Landau damping at present parameters. 

\item The longitudinal distribution that describes the
Recycler bunches is a {\em tanh} distribution. This is found to be
in very good agreement with observations made on separate days and
with different beam parameters. This distribution was used to find
the intensity at which the coherent dipole frequency is at the 
edge of the incoherent frequency distribution in the presence of space
charge. This threshold for the loss of Landau damping is about two 
orders of magnitude above present intensities.

\item An experiment to determine the impact of a local
instability possibly induced at shorter bunch lengths was discussed.
There are at least two competing mechanisms operating in this regime.
On the one hand, at these bunch lengths the extremum of the 
incoherent frequency lies within the bunch frequency distribution
and a local instability may develop. On the other hand we found that 
the threshold for the loss of Landau damping increases as the bunch is
shortened with the area preserved. Nevertheless, beam loss was 
observed at shorter bunch
lengths. Possible causes include higher energy spread, resonance with
the Main Injector ramp frequency and perhaps the local instability.

\item A rough estimate of the microwave instability threshold shows 
that in the absence of space charge, the threshold is about three
orders of magnitude higher than present intensities. Table 
\ref{table: thresh} shows the intensity thresholds from different
effects.
\begin{table}
\bec
\btable{|c|c|} \hline
Effect & Intensity threshold \\ \hline
Loss of Landau damping & 4.7$\times 10^{14}$ \\
Bucket height shrinks to 12 MeV & 2.3$\times 10^{15}$ \\
Microwave instability & 2.9$\times 10^{15}$ \\
\hline
\etable
\eec
\caption{Intensity thresholds for different effects, assuming
$T_1=48$ bkts, $T_2=324$ bkts.}
\label{table: thresh}
\end{table}

\eit

In a subsequent paper we will examine the stability diagrams in the
presence of space charge and a resistive impedance. We will also report
on results of simulations with a macro-particle tracking code (ESME)
and a Vlasov code. The Vlasov code will be employed to compare with
the intensity thresholds found in this paper as well as to understand
the bunch behaviour following the onset of instability. 

\bibliography{dipolemodes}

\end{document}